\documentclass[reprint,amsmath,amssymb,aps,showpacs,prx,longbibliography]{revtex4-2}
\usepackage[utf8]{inputenc}

\usepackage{graphicx}
\usepackage{dcolumn}
\usepackage{bm}
\usepackage[colorlinks=true, linkcolor=blue, filecolor=blue, citecolor = blue, urlcolor=blue]{hyperref}

\usepackage{amsmath}
\usepackage{amssymb}
\usepackage{amsthm}
\usepackage{mathrsfs}

\renewcommand{\bf}[1]{\textbf{#1}}
\usepackage{bm}

\usepackage{subcaption}

\newcommand{\BZ}{\textnormal{BZ}}

\newcommand{\bra}[1]{{\langle{#1}\vert}}
\newcommand{\ket}[1]{{\vert{#1}\rangle}}

\newcommand{\Tr}{\mathop{\rm Tr}\nolimits}
\newcommand{\tr}{\mathop{\rm Tr}\nolimits}
\renewcommand{\Re}{\mathop{\rm Re}\nolimits}

\newcommand{\ds}{\displaystyle}

\usepackage{url}

\usepackage{xcolor}

\newcommand{\pr}{\mbox{pr}}

\begin{document}

\title{Interferometric geometry from symmetry-broken Uhlmann gauge group with applications to topological phase transitions}

\author{Hector Silva}
\email{hectorsilva@tecnico.ulisboa.pt}
\affiliation{Departmento de F\'{i}sica, Instituto Superior T\'ecnico, Universidade de Lisboa, Av. Rovisco Pais, 1049-001 Lisboa, Portugal}

\author{Bruno Mera}
\email{bruno.mera@tecnico.ulisboa.pt}
\affiliation{Instituto de Telecomunica\c{c}\~oes, 1049-001 Lisboa, Portugal}
\affiliation{Departmento de Matem\'{a}tica, Instituto Superior T\'ecnico, Universidade de Lisboa, Av. Rovisco Pais, 1049-001 Lisboa, Portugal}
\affiliation{Departmento de F\'{i}sica, Instituto Superior T\'ecnico, Universidade de Lisboa, Av. Rovisco Pais, 1049-001 Lisboa, Portugal}

\author{Nikola Paunkovi\'{c}}
\email{npaunkov@math.tecnico.ulisboa.pt}
\affiliation{Instituto de Telecomunica\c{c}\~oes, 1049-001 Lisboa, Portugal}
\affiliation{Departmento de Matem\'{a}tica, Instituto Superior T\'ecnico, Universidade de Lisboa, Av. Rovisco Pais, 1049-001 Lisboa, Portugal}

\date{\today}

\begin{abstract}
We provide a natural generalization of a Riemannian structure, i.e., a metric, recently introduced by Sj\"{o}qvist for the space of non degenerate density matrices, to the degenerate case, i.e., the case in which the eigenspaces have dimension greater than or equal to 1. We present a physical interpretation of the metric in terms of an interferometric measurement.  We apply this metric, physically interpreted as an interferometric susceptibility, to the study of topological phase transitions at finite temperatures for band insulators. We compare the behaviors of this susceptibility and the one coming from the well-known Bures metric, showing them to be dramatically different. While both infer zero temperature phase transitions, only the former predicts finite temperature phase transitions as well. The difference in behaviors can be traced back to a symmetry breaking mechanism, akin to Landau-Ginzburg theory, by which the Uhlmann gauge group is broken down to a subgroup determined by the type of the system's density matrix (i.e., the ranks of its spectral projectors).
\end{abstract}

\maketitle


\section{Introduction}
\label{sec: Introduction}

Recent advances in information geometry have provided new methods for studying quantum matter and describing macroscopic critical phenomena based on quantum effects. Topological phases of matter are described in terms of \emph{global} topological invariants that are robust against continuous perturbations of the system. An example of these invariants is the Thouless-Kohmoto-Nightingale-den Nijs (TKNN) invariant, mathematically a Chern number associated with the vector bundle of occupied Bloch states over the Brillouin zone. This invariant captures topological phases of matter that could not be understood previously, such as the case of the anomalous Hall insulator~\cite{hal:88}, which falls into the class of Chern insulators. The classification of topological phases of gapped free fermions is encoded in the so-called periodic table of topological insulators and superconductors~\cite{kit:09}. However, by now we know that these phases of matter were just the tip of an iceberg (see~\cite{shi:sat:14, gon:ash:18, roy:har:17, sch:18}). The theory underlying topological phases constitutes a change of paradigm with respect to the Landau theory of phase transitions~\cite{lan:37}. The latter is described by means of a \emph{local} order parameter, within the framework of the \emph{symmetry-breaking} mechanism.

One can study phases of matter and the associated phase transitions (in particular topological ones) through a Riemannian metric on the space of quantum states. One such commonly used structure is based on the notion of fidelity, which is an information theoretical quantity that measures the distinguishability between quantum states. It has been widely used in the study of phase transitions~\cite{zan:pau:06,pau:sac:nog:vie:dug:08, pau:vie:08, zan:ven:cam:gio:07,zan:goo:coz:07,ven:zan:07,you:wai:gu:07, car:spa:val:18, oza:gol:18, leo:val:spa:car:19, car:val:spag:20}, since its non-analytic behavior signals phase transitions. 

Note that the mentioned topological invariants, being functions of the Hamiltonian only and not the temperature, characterize topological features at zero temperature. Therefore, it is crucial to understand the effect of temperature on topological phase transitions, especially with regard to applications to quantum computers, such as those involving Majorana modes in topological superconductors~\cite{kit:01}. To approach this problem, the fidelity, the associated Bures metric, and, in addition, the Uhlmann connection, the generalization of the Berry connection to the case of mixed states, have been probed for systems that exhibit zero-temperature symmetry-protected topological phases~\cite{mer:vla:pau:vi:17,mer:vla:pau:vie:17:qw,ami:mer:vla:pau:vie:18,sac:mer:pau:19,ami:mer:pau:vie:19}.

Within the context of dynamical phase transitions, occurring when one performs a quench on a system, the information geometric methods based on state distinguishability were applied~\cite{bha:ban:dut:17,mer:vla:pau:vie:viy:18,ban:dut:20a,ban:dut:20b}. In particular, for finite temperature studies, besides the standard notion of fidelity induced Loschmidt echo, a notion of \emph{interferometric Loschmidt echo} based on the interferometric phase introduced by Sj\"{o}qvist \emph{et al.} ~\cite{sjo:00}, was also considered. With regard to the associated infinitesimal counterparts, i.e., Riemannian metrics, their behavior is significantly different. 

For two-band Chern insulators, the fidelity susceptibility, one of the components of the Bures metric, was considered in detail in Ref.~\cite{ami:mer:vla:pau:vie:18}. In particular, it was rigorously proven that the thermodynamic and zero temperature limits do not commute -- the Bures metric is regular in the thermodynamic limit as one approaches the zero temperature limit.

In this paper, we provide, through what is called the Ehresmann connection, a natural generalization of a Riemannian structure over the space of non degenerate density matrices, introduced by Sj\"{o}qvist \cite{sjo:20},  to the degenerate case. Our natural construction reveals a symmetry breaking mechanism by reducing the gauge group of the Uhlmann principal bundle~\cite{uhl:86} to a smaller subgroup preserving the \emph{type} of the density matrix, i.e., the ranks of its spectral projectors (see Sec.~\ref{sec:sjoqvist metric} for details). This symmetry breaking mechanism explains the natural enhanced distinguishability provided by the interferometric Riemannian metric. Introducing the notion of a generalized purification, we naturally generalize Sj\"{o}qvist's result to the case of degenerate density matrices, see Sec.~\ref{sec:distance_measures_and_riemannian_metrics}. In Sec.~\ref{sec:interferometric_measurement}, we discuss an interferometric measurement probing the Riemannian metric derived. In Sec.~\ref{sec:interferometric metric in the context of band insulators}, we apply the derived metric to study finite temperature phase transitions in the context of band insulators. We present results for this metric in the case of the massive Dirac model, a Chern insulator, in two spatial dimensions and compare them with those obtained using the Bures metric. Our analysis of equilibrium phase transitions is shown to be consistent with the previous study of dynamical phase transitions: The interferometric metric is more sensitive to the change in the parameters than the Bures one. Finally, we present conclusions in Sec.~\ref{sec: conclusions}.


%

\section{The geometry of the Sj\"{o}qvist metric and natural generalizations to degenerate cases}
\label{sec:sjoqvist metric}

Consider a quantum system with the corresponding $n$-dimensional Hilbert space $\mathcal H$. Its general mixed state (density matrix) $\rho$ can be, using the spectral decomposition, written as
\begin{equation}
 \label{eq:rho}
\rho = \sum_{i=0}^k p_i P_i,
\end{equation}
where the real eigenvalues satisfy $p_0=0$ and $(i\neq j \Rightarrow p_i \neq p_j)$, while the orthogonal projectors satisfy $(i>0 \Rightarrow \Tr P_i \equiv r_i > 0)$, and $\sum_{i=1}^k r_i = r$. We call $r\in \{ 1, \dots , n \}$ the {\em rank} of the state. Note that we do not require for the kernel of $\rho$ to be nontrivial (i.e., $r_0 \equiv \Tr P_0 \geq 0$), while all other eigenspaces, $\mathcal {H}_i$, are at least one-dimensional (such that $\mathcal {H} = \oplus_{i=0}^k \mathcal {H}_i$). We call the $k$-tuple $\tau \equiv (r_1, r_2, \dots r_k) \in \mathcal T$, with $k \in \{1, \dots, n\}$ and $(1\leq r_1 \leq r_2 \leq \dots \leq r_k)$, the {\em type} of the state $\rho$, where $\mathcal T$ is the set of all possible types. Note that as a consequence of the normalization of density matrices we have the additional constraint 
\begin{align}\label{eq:normalization}
\sum_{i=1}^{k}r_ip_i=1.
\end{align}

Consider the set of all density operators of type $\tau$, denoted by $B_\tau$. The union, over the types $\tau\in \mathcal{T}$, of all sets $B_\tau$ forms the set of all possible states of a given system,
\begin{align}
\label{eq:all_states}
B&= \bigcup_{\tau\in\mathcal T} B_\tau   \\
&= \{\rho \in \mathcal{H}\otimes\mathcal{H}^\ast: \rho^\dag = \rho \ \mbox{and} \ \rho \geq 0 \ \mbox{and} \  \Tr  \rho = 1 \}. \nonumber
\end{align}

We would like to analyze the geometry of the  $B_\tau$'s, and see whether it is possible to induce a Riemannian metric on them along the lines of the metric introduced by Sj\"{o}qvist~\cite{sjo:20}, for the case of type $\tau=(1,1,\dots ,1)$, for some $r=k$. We will do so by introducing gauge invariant Riemannian metrics and associated Ehresmann connections in suitably chosen principal bundles $P_\tau$ with corresponding base spaces $B_{\tau}$. Observe that every state $\rho$ is completely specified in terms of its {\em ``classical part''}, the vector of probabilities $\mathbf{\sqrt{p}} = (\sqrt{p_1}, \sqrt{p_2}, \dots , \sqrt{p_k})$ satisfying the normalization constraint \eqref{eq:normalization}, and its {\em ``quantum part''}, the mutually orthogonal projectors $P_1, P_2, \dots , P_k$ (note that $P_0$ is then determined unambiguously, $P_0=I-\sum_{i=1}^{k}P_i$), which we compactly denote by $\bf{P}=(P_1,P_2,\dots,P_k)$. We will explore a particular gauge degree of freedom in describing the quantum part in our construction. Namely, each eigenspace projector $P_i$ is uniquely specified by an orthonormal basis $\beta_i = \{ \ket{e_{i,j}} : j = 1, \dots r_i\}$. However, the basis $\beta_i$ itself is not uniquely determined by $P_i$. Indeed, every basis $U\beta_i =\{ U\ket{e_{i,j}} :j=1,...,r_i \}$ with $U$ being a unitary that acts non-trivially only on the image of $P_i$, the subspace~$\mathcal{H}_i$, defines the same projector $P_i$. 

We then define (the total space of) a principal bundle $P_\tau$ as the set of all $k$-tuples of pairs $p_{\tau}=\big((p_i,\beta_i)\big)_{i=1}^k$, such that $(\sqrt{\bf{p}},\bf{P})$ give rise to well-defined type $\tau$ density operators (observe that  $p_i\neq p_j$ for all $i\neq j$). This space comes equipped with an obvious projection to the base space $B_\tau$ and is given by
\begin{equation}
\label{eq:projection}
\pi_\tau (p_\tau) \equiv \sum_{i=1}^k p_i P_i = \rho,
\end{equation}
with the fibers being isomorphic to the product of the corresponding unitary groups in the type $\tau$,
\begin{equation}
\label{eq:fibre}
 G_\tau \equiv \prod_{i=1}^k \mbox{U}(r_i).
\end{equation}
The group $G_{\tau}$ acts on the right in the obvious way, for $U_i\in \mbox{U}(r_i)$, we write $U_i=[(U_i)^{j'}_{\;j}]_{1\leq j,j'\leq r_i}\in \mbox{U}(r_i)$ and then $\beta_i\cdot U_i$ is  given by
\begin{align}
\ket{e_{i,j}}\mapsto \sum_{j'=1}^{r_i}\ket{e_{i,j'}}(U_i)^{j'}_{\;j},\ j=1,...,r_i.
\end{align}

By introducing \emph{generalized amplitudes} $w_i \in \mathbb{C}^{n\times r_i}$ as matrices whose columns are vectors $\ket{e_{i,j}}\in\mathbb{C}^n$, $j=1,...,r_i$, i.e., $w_i \equiv \big( \ket{e_{i,1}} \ \ket{e_{i,2}} \dots \ket{e_{i,r_i}} \big)$, $i=1,...,k$, we can see $P_\tau$ as
\begin{align}
\label{eq:principal_bundle_2}
P_\tau =& \{ \big((p_i,w_i)\big)_{i=1}^k : \sum_{i=1}^{k} p_i\;w_iw_i^\dagger \in B_{\tau} \nonumber\\
&\text{ and } w_i^\dagger w_i=I_{r_i}, \text{ for all } i=1,...,k,\\
&\text{ and } p_i\neq p_j, \text{ for all } i\neq j\}, \nonumber
\end{align}
and the right action of the gauge group is given by $w_i \mapsto w_i\cdot U_i$, with $U_i \in \mbox{U}(r_i)$. With this notation, we finally introduce a suitable ``Hermitian form'' (note that it is not a scalar product, as $P_\tau$ is not a linear space), which will define horizontal subspaces, by the formula
\begin{align}
\label{eq:hermitian_form}
\langle p_\tau , p_\tau^\prime \rangle_\tau &\equiv \sum_{i=1}^k \sqrt{p_i p_i^\prime} \Tr(w_i^\dag w_i^\prime) \nonumber \\
&= \sum_{i=1}^k \Tr[(\sqrt{p_i}w_i^\dag) (\sqrt{p_i^\prime}w_i^\prime)].
\end{align}
Observe that it is clear that this pairing arises from the restriction of the usual Hermitian inner product in $\bigoplus_{i=1}^{k}\mathbb{C}^{n\times r_i}\cong \mathbb{C}^{n\times r}$. 

Additionally, this allows for a convenient comparison with the Uhlmann principal bundle
\begin{align}
\label{eq:uhlmann_bundle}
 P_{r}^{\text{Uh}} =& \{ w\in\mathbb{C}^{n\times r} : \pi(w)\equiv ww^\dag  = \rho\in B, \nonumber\\
 &\text{ with } \text{rank}(\rho)=r \},
\end{align}
where the typical fiber is $\mbox{U}(r) \subset \mathbb{C}^{r\times r}$, whose elements act from the right ($w \mapsto w\cdot U$), and the Hermitian form, induced by the Hilbert-Schmidt scalar product on the space of linear operators from $\mathbb{C}^{r\times r}$, is
\begin{equation}
\label{eq:hilbert-schmidt}
  \langle w , w^\prime \rangle = \Tr(w^\dag w^\prime).
\end{equation}
Note that the base space for the Uhlmann bundle is the set of density matrices with rank $r$, which is the union of all $B_{\tau}$ sharing the same rank. Observe that for one such $\tau$, $P_{\tau}$ can be identified as a subset of $P_{r}^{\text{Uh}}$. This follows from the map
\begin{align}
\label{eq: embedding of ptau}
P_{\tau} \ni \left((p_i,w_i)\right)_{i=1}^{k}\mapsto (\sqrt{p_1}w_1,....,\sqrt{p_k}w_k) \in \bigoplus_{i=1}^{k}\mathbb{C}^{n\times r_i},
\end{align}
being an embedding of $P_{\tau}$. Moreover, once we identify $\bigoplus_{i=1}^{k}\mathbb{C}^{n\times r_i}\cong \mathbb{C}^{n\times r}$, the image sits precisely in $P_{r}^{\text{Uh}}$. In other words $P_{\tau}\subset P_{r}^{\text{Uh}}$ and also $\pi_{\tau}$ equals the restriction of the projection of the Uhlmann bundle to $P_{\tau}$ ($p_i\neq p_j$, for all $i\neq j$, guarantees this), the image being precisely $B_{\tau}$. We remark that the gauge group of the Uhlmann bundle is far larger than the one for the principal bundle $P_{\tau}\to B_{\tau}$. By passing to a preferred type, we performed a symmetry breaking operation from $\mbox{U}(r)$ to $G_{\tau}=\prod_{i=1}^{k}\mbox{U}(r_i)\subset \mbox{U}(r)$. This is another way to see why interferometric-like quantities, such as the interferometric Loschmidt echo, in certain applications develop non-analyticities, while the ones based on the fidelity do not (see for example~\cite{mer:vla:pau:vie:viy:18} and the references therein): The former have smaller space to ``go through'', while the latter can, following the ``broader'' Uhlmann connection, instead of the interferometric ones, avoid possible sources of non-analyticities.

\section{Distance measures and Riemannian metrics}
\label{sec:distance_measures_and_riemannian_metrics}
Consider now two points, $p_{\tau}=\left((p_i,w_i)\right)_{i=1}^{k}$ and $q_{\tau} = \left( \left( q_i, v_i\right) \right)_{i=1}^k \in P_{\tau}$. By making use of Eq.~\eqref{eq:hermitian_form} one can define a distance between elements $p_\tau$ and $q_\tau$ in the total space of the principal bundle given by

\begin{align}
    d_\tau^2(p_{\tau},q_{\tau} ) &= 2 \Big( 1 -\Re\left( \langle p_{\tau} , q_{\tau} \rangle_\tau\right)  \Big) \nonumber\\
    &= 2 \left( 1 - \sum_{i=1}^k \sqrt{p_i q_i} \ \Re \left(\Tr(w_i^{\dagger} v_i)\right)\right).
\end{align}
The fact that $d_\tau$ is a distance follows from the fact that it is the restriction of the usual distance in $\oplus_{i=1}^{k}\mathbb{C}^{n\times r_i}$, where we see $P_{\tau}$ as a subset of this space through the map of Eq.~\eqref{eq: embedding of ptau}.
One can use this distance to define a distance on $B_{\tau}$, through the formula:
\begin{eqnarray}
	\label{eq:distance_btau}
d^2_{I}(\rho,\sigma)= \inf \{ && d_\tau^2(p_{\tau},q_{\tau}): \pi(p_{\tau})=\rho \\ && \text{ and }\pi(q_{\tau})=\sigma, \text{ for } p_{\tau},q_{\tau}\in P_{\tau} \}.\nonumber
\end{eqnarray}
The associated infinitesimal counterparts of the distances defined above are Riemannian metrics on $P_{\tau}$ and $B_{\tau}$, respectively. The Riemannian metric on $P_{\tau}$, which is gauge invariant, allows for the definition of what is called an Ehresmann connection over $P_{\tau}$ and this, in turn, defines a metric downstairs over the base space $B_{\tau}$.

Another way to see that $d^2_\tau (p_\tau , q_\tau )$ is indeed a metric is through what we call {\em``generalized purifications''}. Let us introduce  {\em ``ancilla''} amplitudes $\mbox{\cal{w}}_i\in \mathbb{C}^{k\times 1}$, with $i = 1,2,\dots
 k$, such that $\mbox{\cal{w}}_i \mbox{\cal{w}}_i^\dag = \mbox{P}_i \in \mathbb{C}^{n\times n}$ are {\em fixed} orthogonal projectors of rank $1$ (i.,e., $\mbox{P}_i$ do not depend on the choice of the state), satisfying $\mbox{P}_i\mbox{P}_j = \delta_{ij}I_k$ and $\sum_{i=1}^k \mbox{P}_i = I_k$. Define a generalized purification of state $\rho$, associated with the corresponding $p_\tau$, as
 \begin{equation}
  \ket{p_\tau} = \sum_{i=1}^k \sqrt{p_i} w_i \otimes \mbox{\cal{w}}_i.	
 \end{equation}
Then, we have that the scalar product between $\ket{p_\tau}$ and $\ket{q_\tau}$, induced by the Hilbert-Schmidt scalar product in the corresponding factor spaces, is
\begin{equation}
\begin{array}{rcl}
\langle p_\tau , q_\tau \rangle \!\!\! & = & \!\!\! \ds \sum_{i,j=1}^k \sqrt{p_iq_j} \langle w_i , v_j\rangle\langle \mbox{\cal{w}}_i , \mbox{\cal{w}}_j\rangle \\
\!\!\! & = & \!\!\! \ds \sum_{i=1}^k \sqrt{p_iq_i} \langle w_i , v_i \rangle \\
\!\!\! & = & \!\!\! \ds \sum_{i=1}^k \sqrt{p_iq_i} \Tr(w_i^\dag v_i) \\[0.5cm] 
\!\!\! & = & \!\!\!
\langle p_\tau , q_\tau \rangle_\tau ,
\end{array}
\end{equation}
where the second equality is because $\mbox{\cal{w}}_i$ and $\mbox{\cal{w}}_j$ are orthogonal for $i \neq j$. Thus the distance $d_\tau(p_{\tau},q_{\tau} )$ is nothing but the standard Hilbert-Schmidt distance between the generalized purifications $\ket{p_\tau}$ and $\ket{q_\tau}$.

As in Eq.~\eqref{eq:principal_bundle_2}, if we take the $w_i$'s as (row) vectors $\ket{w_i}=\bigg[\ket {e_{i,1}} \ \ket {e_{i,2}} \ \dots \ \ket {e_{i,r_i}}\bigg]$ whose entries are (column) vectors $\ket {e_{i,j}}$, one can by analogy generalize the quantum part of the metric for the non-degenerate case, the so-called \emph{``interferometric metric''}, which has $r_i=1$, $i=1,...,k$, 
\begin{equation}
\begin{array}{rcl}
	g_{I}^{\text{Q}} \!\!\! & = & \!\!\! \ds \sum_{i=1}^k p_i \bra{dw_i} (I_n - w_iw_i^\dag) \ket{dw_i} \\
	\!\!\! & = & \!\!\! \ds \sum_{i=1}^k p_i \bra{de_{i,1}} (I_n - \ket{e_{i,1}}\bra{e_{i,1}}) \ket{de_{i,1}},
	\end{array}
\end{equation}
to the degenerate case, in which $\mbox{U}(1)$ degree of freedom of each $w_i = \ket{e_i}$ is replaced by the $\mbox{U}(r_i)$ degree of freedom of each $w_i = \bigg[\ket {e_{i,1}} \ \ket {e_{i,2}} \ \dots \ \ket {e_{i,r_i}}\bigg]$,
\begin{equation}
\begin{array}{rcl}
	g_{I}^{\text{Q}} \!\!\! & = & \!\!\! \ds \sum_{i=1}^k p_i \bra{dw_i} \big(I_n - w_iw_i^\dag \big) \ket{dw_i} \\
	\!\!\! & = & \!\!\! \ds \sum_{i=1}^k p_i \bra{dw_i} \Big[I_n - \big(\sum_{j=1}^{r_i} \ket{e_{i,j}}\bra{e_{i,j}}\big) \Big] \ket{dw_i} \\
	\!\!\! & = & \!\!\! \ds \sum_{i=1}^k p_i \bra{dw_i} \big( I_n - P_i \big) \ket{dw_i},
	\end{array}
\end{equation}
with $\ket{dw_i}=\bigg[\ket {de_{i,1}} \ \ket {de_{i,2}} \ \dots \ \ket {de_{i,r_i}}\bigg]$, $i=1,...,k$.
Indeed, in Appendix~\ref{sec:induced_metrics} we prove that this intuitive generalization is the correct result describing the infinitesimal counterpart of the distance in 
Eq.~\eqref{eq:distance_btau}.

\section{Interferometric measurement interpretation}
\label{sec:interferometric_measurement}

\begin{figure}[h]
\centering
\includegraphics[width=0.75\linewidth]{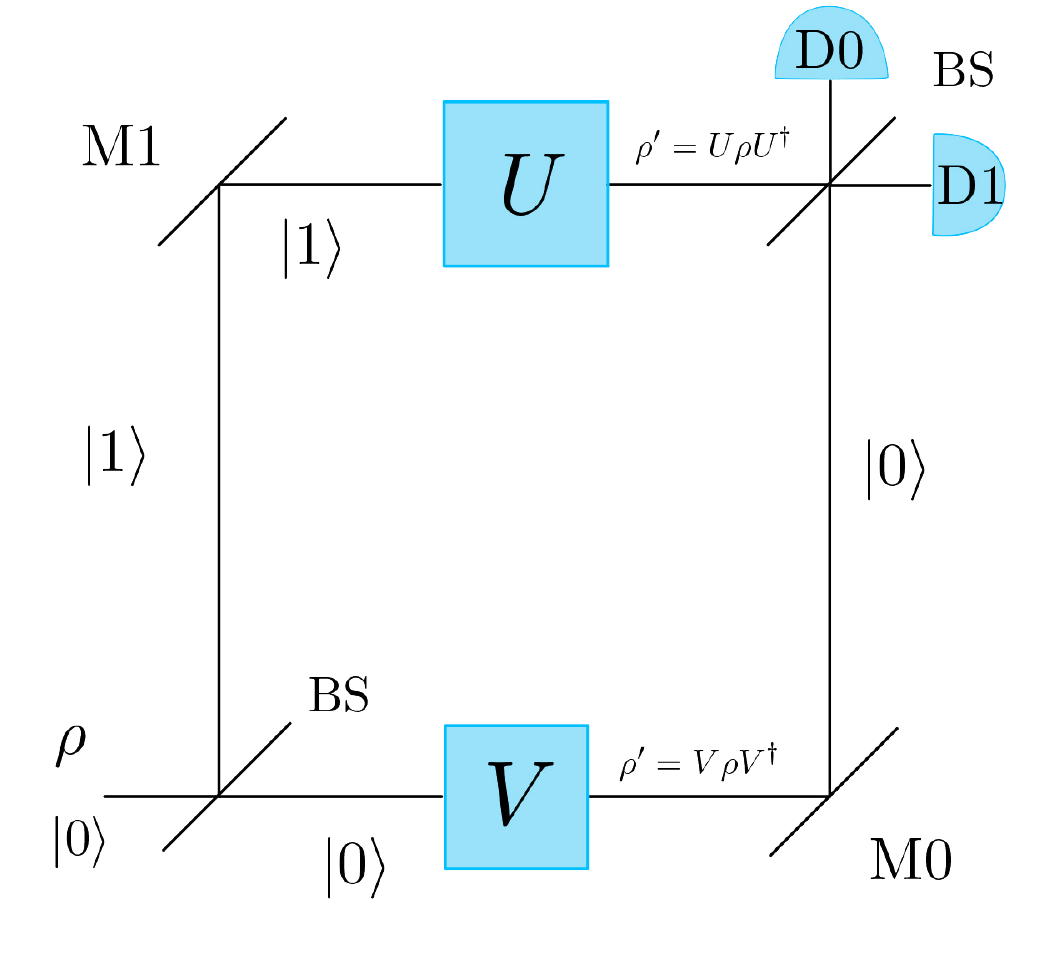}
\caption{Interferometric measurement to probe the generalized metric $g_{I}$. $\mbox{BS}$ represents beam-splitter, $\mbox{M}_0$ and $\mbox{M}_1$ mirrors in arms $0$ and $1$, respectively, and analogously for detectors $\mbox{D}_0$ and $\mbox{D}_1$.}
\label{fig:mach-zehnder}
\end{figure}

Consider the following experiment depicted in FIG~\ref{fig:mach-zehnder}. A particle is entering the Mach-Zehnder interferometer from the input arm 0, given by the state $\ket 0$, with its internal degree of freedom in a mixed state $\rho$. Both the input and the output beam-splitters are balanced, described by the same unitary matrix, say, the one given by $\ket{0} \rightarrow (\ket 0 + i\ket 1)/\sqrt{2}$. In arm 0 a unitary $V= \sum_{i=0}^k P_i VP_i$ is applied to the internal degree of freedom, that is, $V$ is the most general unitary that commutes with $\rho$. In arm 1 a  unitary $U = U(\delta t) \in \mbox{U}(n)$ is applied for a time period $\delta t$, changing the state of the internal degree of freedom to $\rho^\prime = U \rho U^\dag$. The particle is detected at detectors D0 and D1, with the corresponding probabilities $\mbox{pr}_0$ and $\mbox{pr}_1$. In our case, we have that $\mbox{pr}_1 \leq \mbox{pr}_0$, and for $U=V$ we have full constructive interference at the output arm $0$, giving $\mbox{pr}_0=1$. In general, we have that
\begin{equation}
	\mbox{pr}_1^{\mbox{max}} = \max_{\{V_i\}} ( \mbox{pr}_1 ) = 1 - \frac{1}{4}d^2_{I}(\rho,\rho+\delta \rho),
\end{equation}
where $d^2_{I}(\rho,\rho+\delta \rho)\approx g_{I}(\dot{\rho},\dot{\rho})\delta t^2$ is the ``infinitesimal'' distance between $\rho$ and $\rho^\prime = \rho +\delta \rho$, where $\delta \rho= \dot{\rho}\delta t$ (see Appendix~\ref{sec:interferometric_probability} for a detailed proof). Note that in the case of the Hadamard matrix, given by $\ket{\ell} \rightarrow (\ket 0 + (-1)^\ell\ket 1)/\sqrt{2}$, with $\ell \in \{ 0,1\}$, the roles of arms 0 and 1 are exchanged.

\section{Interferometric metric in the context of band insulators}
\label{sec:interferometric metric in the context of band insulators}
Suppose we have a family of band insulators with two bands described by the Hamiltonian
\begin{align}
\label{eq:band insulator}
\mathcal{H}(M)=\int_{\BZ^d} \frac{d^dk}{(2\pi)^d}\; \psi^\dagger_{\bf{k}}d^\mu(\bf{k};M)\sigma_\mu\psi_{\bf{k}},
\end{align}
parametrized by $M$ ($M$ can be some intrinsic parameter, such as the hopping), where $\sigma_\mu$, $\mu=1,2,3$, are the Pauli matrices, $\bf{k}$ is the crystalline momentum in a $d$-dimensional Brillouin zone $\BZ^d$, with $d=1,2,3$, and $\Psi^\dagger_{\bf{k}}$ is an array of two creation operators for fermions at momentum $\bf{k}$. We assume that the system is gapped for generic values of $M$, meaning that the vector $d=(d^1,d^2,d^3)$ is non-vanishing as a function of $\bf{k}$. For a certain value of $M_c$, we assume that the vector has isolated zeros. This assumption is generically correct for the $d=1,2$ momenta coordinates plus the mass $M$, as one needs to tune three parameters for a Hermitian matrix to have two eigenvalues cross. 

The pullback of the interferometric metric that we have described in Sec.~\ref{sec:distance_measures_and_riemannian_metrics},
\begin{align}
g=\frac{1}{4}\sum_i r_i \frac{dp_i^2}{p_i}+\sum_{i}p_i \Tr\left(P_idP_idP_i\right), 
\end{align}
with $\rho=\sum_i p_i P_i$ and $\Tr P_i=r_i$, by the map induced by the Gibbs state
\begin{align}
M \mapsto \rho(M)=Z^{-1}\exp(-\beta \mathcal{H}(M)),
\end{align}
where $Z$ is the partition function, is given by
\begin{align}
ds^2&=\frac{1}{4}\int_{\BZ^d}\frac{d^d k}{(2\pi)^d}\Big[ \frac{1}{\cosh(\beta E)+1}\\
&\times \left(\beta^2\left(\frac{\partial E}{\partial M}\right)^2 + \cosh(\beta E)\delta_{\mu\nu}\frac{\partial n^\mu}{\partial M}\frac{\partial n^\nu}{\partial M}\right)\Big]dM^2, \nonumber
\end{align}
where we omitted the obvious dependence on $\bf{k}$ and $M$ of the quantities $E$ and $n^\mu$. We provide a technical derivation of this result in Appendix~\ref{sec: pullback of interferometric metric to parameter space}. This result should be compared with the pullback of the Bures metric for $d=2$, which yields (see Ref.~\cite{ami:mer:vla:pau:vie:18})
\begin{align}
g_{\text{Bures}}=\frac{1}{4}\int_{\BZ^d}&\frac{d^d k}{(2\pi)^d} \Big[ \frac{1}{\cosh(\beta E)+1}\beta^2\left(\frac{\partial E}{\partial M}\right)^2 \\
&+ \frac{\cosh(\beta E)-1}{\cosh(\beta E)}\delta_{\mu\nu}\frac{\partial n^\mu}{\partial M}\frac{\partial n^\nu}{\partial M}\Big]dM^2. \nonumber
\end{align}
The two expressions have dramatically different behaviors, when it comes to taking the zero temperature limit. 

Naively, one would say that both yield the pullback of the Fubini-Study metric, which is the pure-state metric,
\begin{align}
g_0=\frac{1}{4}\int_{\BZ^d} \frac{d^d k}{(2\pi)^d} \delta_{\mu\nu} \frac{\partial n^\mu}{\partial M}\frac{\partial n^\nu}{\partial M} dM^2.
\end{align}
Note that for gapless points the vector $n$ is not defined and the expression for $g_0$ becomes (potentially) singular. However, due to the gapless points, the integrands must be carefully analyzed in the neighborhoods of these points, as the singularities can be avoided in some cases. In fact, it was shown that if the gapless points are isolated in momentum space, then an expansion near these points of the integrand function yields a regular result~\cite{ami:mer:vla:pau:vie:18}. Namely, because of the inequality
\begin{align}
\!\!\!\frac{1}{2}\frac{1}{\cosh(x)}\!<\!\frac{1}{\cosh(x)+1} \!<\! \frac{1}{\cosh(x)}, \text{ for all } x\in\mathbb{R},
\end{align}
we can write,
\begin{align}
&\frac{1}{\cosh(\beta E)\!+\!1}\beta^2\!\!\left(\!\!\frac{\partial E}{\partial M}\!\!\right)^2 \!\!\!+\! \frac{\cosh(\beta E)\!-\!1}{\cosh(\beta E)}\delta_{\mu\nu}\frac{\partial n^\mu}{\partial M}\frac{\partial n^\nu}{\partial M} \\
&<\frac{1}{\cosh(\beta E)}\left[\beta^2\left(\frac{\partial E}{\partial M}\right)^2 +\left(\cosh(\beta E)-1\right)\delta_{\mu\nu}\frac{\partial n^\mu}{\partial M}\frac{\partial n^\nu}{\partial M}\right]. \nonumber
\end{align}
Expansion for small $\beta E$ yields that up to $\mbox{O}\left((\beta E)^4\right)$ the integrand is upper bounded by
\begin{align}
\frac{\beta^2}{\cosh(\beta E)}\delta_{\mu\nu}\frac{\partial d^\mu}{\partial M}\frac{\partial d^\nu}{\partial M},
\end{align} 
which is regular in the limit $\beta\to \infty$. Hence, the potential singularities arising from the gapless region are regularized by the Bures prescription. However, in the case of the interferometric metric, considering the integrand
\begin{align}
\frac{1}{\cosh(\beta E)+1}&\!\Big(\!\beta^2 \!\!\left(\!\frac{\partial E}{\partial M}\!\right)^2 \!\!+\! \cosh(\beta E)\delta_{\mu\nu}\frac{\partial n^\mu}{\partial M}\frac{\partial n^\nu}{\partial M}\Big),
\end{align}
near $E=0$ gives us
\begin{align}
\frac{1}{\cosh(\beta E)+1}\Big[ &\beta^2\left(\frac{\partial E}{\partial M}\right)^2 + (1+\frac{1}{2}\beta^2 E^2)\delta_{\mu\nu}\frac{\partial n^\mu}{\partial M}\frac{\partial n^\nu}{\partial M}\nonumber\\
&+\mbox{O}\left((\beta E)^4\right)\Big].
\end{align}
In this case, we cannot get rid of the singular factor 
\begin{align}
\delta_{\mu\nu}\frac{\partial n^\mu}{\partial M}\frac{\partial n^\nu}{\partial M},
\end{align}
which appears once in the second term without the regularizing coefficient $\beta^2 E^2$ which above allowed for the identification of the regular quantity
\begin{align}
\!\!\! \beta^2\left(\frac{\partial E}{\partial M}\right)^2 +\beta^2E^2 \delta_{\mu\nu}\frac{\partial n^\mu}{\partial M}\frac{\partial n^\nu}{\partial M}= \beta^2\delta_{\mu\nu}\frac{\partial d^\mu}{\partial M}\frac{\partial d^\nu}{\partial M}.
\end{align}
This implies that the limit $\beta\to\infty$ yields singular behavior for $g$, provided the same happens with $g_0$, but not the other way around, that is, singular behavior on the finite temperature metric does not imply zero temperature singular behavior. In other words, while in the case of the Bures metric the thermodynamic and the zero temperature limits did not commute, in the interferometric case they do, because the singular behavior of the gapless points is recovered, as one considers a small neighborhood of these points and takes the zero temperature limit.
In the following, we will consider the massive Dirac model to illustrate the different behaviors of the two metrics.
\subsection{Massive Dirac model}
\label{subsec:massive dirac model}
\begin{figure}
\begin{subfigure}[b]{0.4\textwidth}
\centering
\includegraphics[scale=0.5]{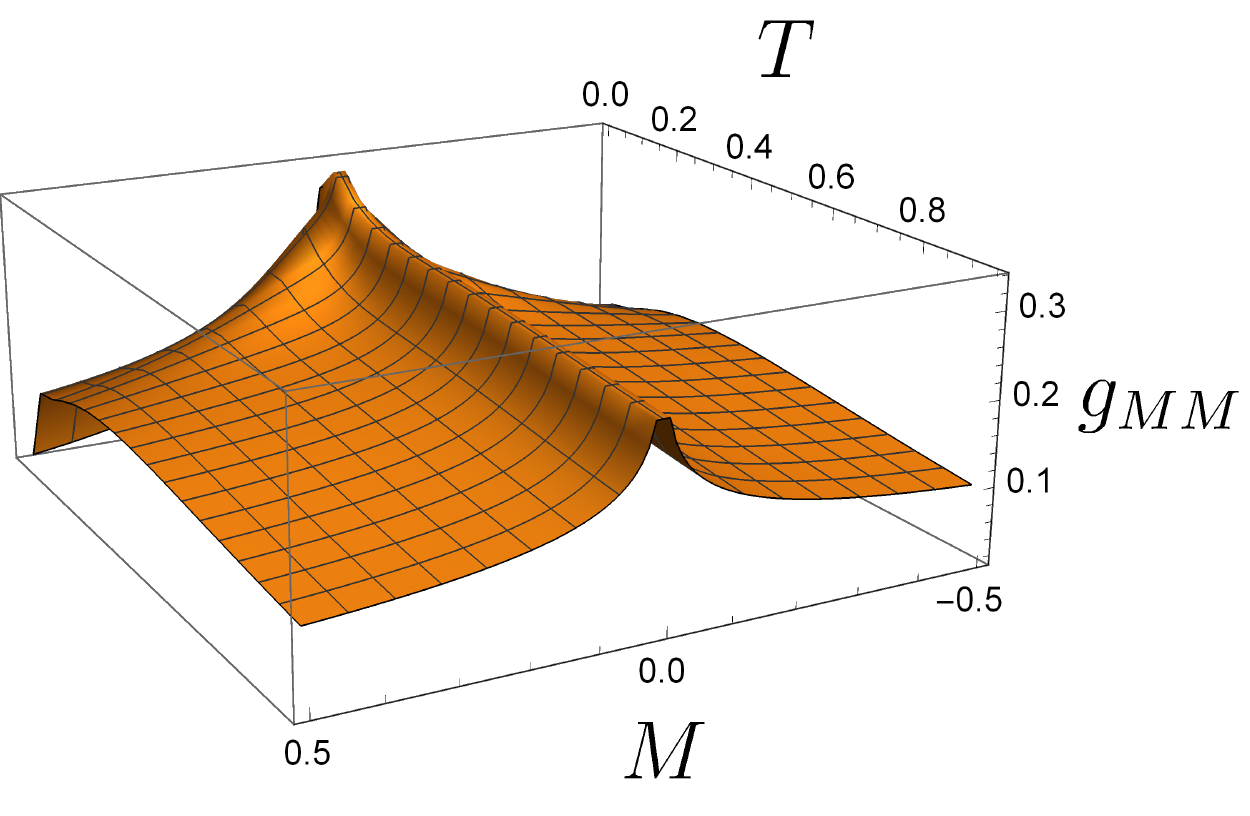}
\caption{Interferometric metric for the massive Dirac model -- the topological phase transition is captured for all temperatures.}
\label{fig:interferometric}
\end{subfigure}
\begin{subfigure}[b]{0.4\textwidth}
\centering
\includegraphics[scale=0.5]{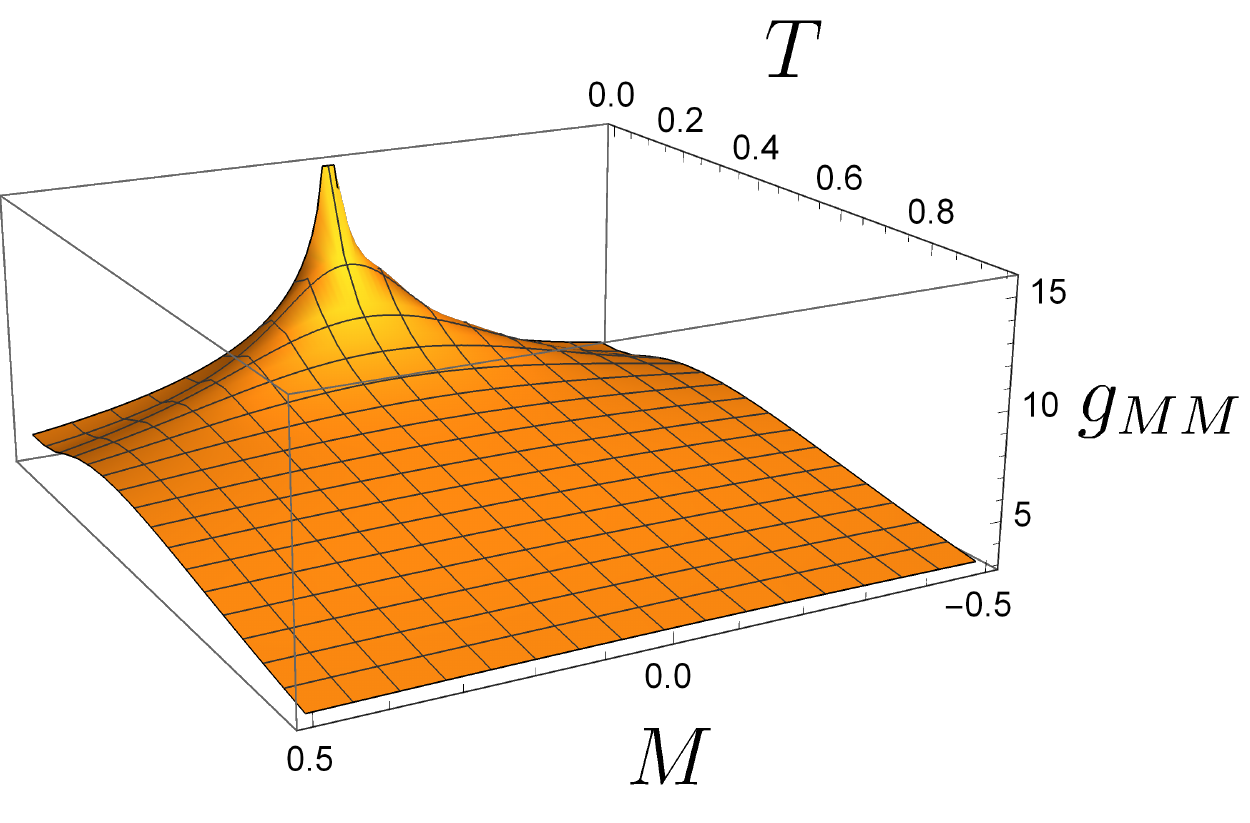}
\caption{Bures metric for the massive Dirac model -- the topological phase transition is captured only at zero temperature.}
\label{fig:bures}
\end{subfigure}
\caption{The different behavior of the metrics with temperature $T$ and the parameter $M$ driving the topological phase transition.}
\end{figure}

We consider the massive Dirac model, a band insulator in two spatial dimensions, described by Eq.~\eqref{eq:band insulator}, with
\begin{align}
\!\!d(\bf{k};M)&\!=\!\left(\sin(k_x),\sin(k_y),M\!-\!\cos(k_x)\!-\!\cos(k_y)\right),
\end{align}
where $\bf{k}=(k_x,k_y)$ is the quasi-momentum in the two-dimensional Brillouin zone $\BZ^2$ and $M$ is a real parameter. The model exhibits topological phase transitions~\cite{mat:ryu:10}. We will focus on the one occurring at $M=0$, where the Chern number goes from $+1$, for $M\to 0^{-}$, to $-1$, for $M\to 0^{+}$. 
Figure 2 describes the interferometric metric (Fig.~\ref{fig:interferometric}) and the Bures metric (Fig.~\ref{fig:bures}) in the thermodynamic limit.

As argued above, the Bures metric is regular if one considers the thermodynamic limit and then the zero temperature limit. The same does not hold for the interferometric metric. In fact, we can see that the interferometric metric knows about the quantum phase transition taking place at $T=0$ even at finite temperatures. The reason is that in passing from one metric to the other the symmetry was broken, namely $\mbox{U}(r)\rightarrow \prod_{i=1}^{k}\mbox{U}(r_i)$, and therefore there is enhanced distinguishability. Indeed, in the interferometric case, whenever the gap closes, we expect a phase transition, even at finite temperatures, because then there are states which according to a Boltzmann-Gibbs distribution become degenerate in probability; hence the gap closing changes the type of the density matrix involved. Whether such singular behavior of the interferometric metric is indeed observable for macroscopic many-body systems is an open question. While the straightforward implementation of the interferometric experiment described in Sec.~\ref{sec:interferometric_measurement} seems to be, at least technologically, infeasible, as it would require maintaining Schr\"{o}dinger cat-like macroscopic states, possible variations are argued to be able to reveal the singular behavior of the interferometric metric at finite temperatures (see Sec.~V of Ref.~\cite{bar:waw:alt:fle:die:18}).
\section{Conclusions}
\label{sec: conclusions}
In this paper, we have generalized Sj\"{o}qvist's interferometric metric introduced in Ref~\cite{sjo:20}, to the degenerate case. For this purpose, we have introduced generalized amplitudes and purifications. We have analyzed an interpretation of the metric in terms of a suitably generalized interferometric measurement, accommodating for the non-Abelian character of our gauge group, as opposed to the Abelian gauge group used in the non degenerate case. We have applied the induced Riemannian structure, physically interpreted as a susceptibility, to the study of topological phase transitions at finite temperatures for band insulators. To the best of our knowledge, this is the first study of finite-temperature equilibrium phase transitions using interferometric geometry. The inferred critical behavior is very different from that of the Bures metric.  The interferometric metric is more sensitive to the change in parameters than the Bures one, and unlike the latter, in addition to zero temperature phase transitions, infers finite temperature phase transitions as well. This sensitivity can be traced back to a symmetry breaking mechanism, much in the same spirit of the Landau-Ginzburg theory. In our case, by fixing the type of the density matrix considered, a gauge group is broken down to a subgroup.

It would be very interesting to analyze the interferometric curvature, an analog of the usual Berry curvature, generalized to this mixed setting, associated with the Ehresmann connection presented in this paper. Since the curvature is intrinsically related to topological phenomena, this analysis might very well unravel new symmetry protected topological phases in the mixed state case and potentially help refine the classification of topological matter.
It would be also interesting to compare the critical behavior of different many-body systems in terms of interferometric metrics corresponding to different types of density matrices. Recent study of the fidelity susceptibility indicated that its singular behavior around regions of criticality has preferred directions on the parameter space~\cite{ami:mer:pau:vie:20}. Performing a similar analysis for the interferometric critical geometry is another possible line of future research. Finally, probing experimentally the introduced interferometric metrics is a relevant topic of future investigation.


%
\section*{Acknowledgments}
\label{sec: Acknowledgements}

B.M. and N.P. are thankful for the support from SQIG -- Security and Quantum Information Group,  the Instituto de Telecomunica\c{c}\~oes (IT) Research
Unit, Ref. UIDB/50008/2020, funded by Funda\c{c}\~ao para a Ci\^{e}ncia e a Tecnologia (FCT), European funds, namely, H2020 project SPARTA, as well as  projects QuantMining POCI-01-0145-FEDER-031826 and PREDICT PTDC/CCI-CIF/29877/2017. N.P. acknowledges FCT Est\'{i}mulo ao Emprego Cient\'{i}fico grant
no. CEECIND/04594/2017/CP1393/CT0006. 

\appendix
\section{Induced Riemannian metrics}
\label{sec:induced_metrics}
Let us look again at the principal bundle $P_\tau$, for a fixed type $\tau=(r_1,...,r_k)$. In this case, a point in $P_{\tau}$ is given by $p_\tau = \big((p_i,w_i)\big)_{i=1}^k $ and can be equivalently represented as $p_\tau = \big((p_i)_{i=1}^k , (w_i)_{i=1}^k\big)$. With this identification, we can separate $p_{\tau}$ into its ``\emph{classical}'' and ``\emph{quantum}'' parts:
\begin{itemize}
\item[(i)] A classical probability amplitude vector $\sqrt{\bf{p}}=(\sqrt{p_1},...,\sqrt{p_k})$, with $\sum_{i=1}^{k}p_i=1$ and, for each $i\in\{1,...,k\}$, $p_i>0$. Note that the set of all classical probability amplitudes is in fact contained in the $(k-1)$-dimensional sphere and the associated classical Fisher metric is, up to a factor of $1/4$, the usual round metric in the sphere $S^{k-1}$.
\item[(ii)] A quantum part which is a $k$-tuple, i.e., a sequence of matrices $(w_{1},...,w_k)$, each of them identifying a $r_i$-unitary frame in $\mathbb{C}^n$, i.e., $w_i\in \mbox{V}_{r_i}(\mathbb{C}^n)$, where
\begin{align}
V_{r_i}(\mathbb{C}^n) &= \{ w_i \in \mathbb{C}^{n \times r_i} : w_i^\dagger w_i = I_k \} \subset \mathbb{C}^{n \times r_i},\nonumber\\
\ & i=1,...,k,
    \label{eq:stiefel}
\end{align}
commonly known as the \emph{Stiefel} manifold of $r_i$-unitary frames in $\mathbb{C}^n$.
\end{itemize}

Our aim is to compute the Riemannian metric in the base space $B_{\tau}$ for a given type $\tau=(r_1,...,r_k)$. For this purpose, we will first look at the tangent space at a point $p_{\tau}$, which is isomorphic to the direct sum
\begin{equation}
    T_{p_\tau}P_\tau \cong T_{\sqrt{\bf{p}}} S^{k-1}\oplus \left(\bigoplus_{i=1}^{k} \, T_{w_i}V_{r_i}(\mathbb{C}^n) \right).
    \label{eq: tangentbundle}
\end{equation}
This isomorphism follows from the factorization into classical and quantum parts: For every curve in the total space $P_{\tau}$, there will be a tangent vector for each of the curves induced by projection in the different factors of $P_\tau$.

The classical components have no gauge ambiguity. The quantum components, however, have a $\mbox{U}(r_i)$  gauge degree of freedom for each matrix $w_i$, $i=1,...,k$. This gauge ambiguity corresponds to variations along the fibers, as we will mention later on. From a physical standpoint, the exact point in the fiber has no significance, since the matrices $w_i$ will be projected onto the base space, where the projectors $P_i$ are gauge invariant: namely, $w_i$ and $w_i\cdot U$, for $U\in\mbox{U}(r_i)$, give rise to the same projector $P_i=w_iw_i^\dagger=w_iUU^\dagger w_i^\dagger$, for all $i=1,...,k$. Hence, we need to define the horizontal subspaces of the tangent spaces to $P_\tau$, in order to uniquely represent the tangent spaces to the base space upstairs, i.e., in the tangent spaces to $P_\tau$. Mathematically, this notion is referred to as an \emph{Ehresmann connection}, see, for example, Sec.~6.3 of Ref.~\cite{mor:01}.

Before we proceed, let us focus on one of the Stiefel manifolds, say for a fixed $i\in\{1,...,k\}$, $V_{r_i}(\mathbb{C}^n)$. For convenience, we define the projection onto the space of projectors of rank $r_i$, identified with the Grassmannian of $r_i$-planes in $\mathbb{C}^n$, i.e., the manifold of linear subspaces of dimension $r_i$ in $\mathbb{C}^n$, 
\begin{equation}
\begin{array}{c}
\pi_{i}: V_{r_i}(\mathbb{C}^n)\to \mbox{Gr}_{r_i}(\mathbb{C}^n)\\[0.2cm]
 w_i\mapsto P_i=w_iw_i^\dagger.
\end{array}
\end{equation}
Consider a curve in the Stiefel manifold
\begin{equation}
    \gamma_{w_i} :\left[0 , 1\right]\ni t \mapsto \gamma_{w_i}(t) \in V_{r_i}(\mathbb{C}^n)
\end{equation}
subject to the initial conditions $\gamma_{w_i}(0) = w_i $ and $\dfrac{d \gamma_{w_i}}{dt}\Big|_{t=0} = \dot{w}_i\equiv\widetilde{v}$. From the definition of $V_{r_i}(\mathbb{C}^n)$, the tangent spaces are
\begin{equation}
    T_{w_i}V_{r_i}(\mathbb{C}^n) =  \{ \dot{w}_i \in \mathbb{C}^{n \times r_i} : \dot{w}_i^\dagger w_i + w_i^\dagger \dot{w}_i = 0 \}.
    \label{eq:T_space}
\end{equation}
The vertical space at $w_i\in V_{r_i}(\mathbb{C}^n)$ is the set of tangent vectors in $T_{w_i} V_{r_i}(\mathbb{C}^n)$, such that its infinitesimal projection onto the base space is zero, that is
\begin{align}
    &\dfrac{d}{dt}\left( \pi_{i} \left( \gamma_{w_i}(t) \right)  \right)\Big|_{t=0} = 0\nonumber\\ \Leftrightarrow &\dfrac{d}{dt} \left( \gamma_{w_i}(t) \gamma_{w_i}^\dagger(t) \right)\Big|_{t=0} = \dot{w}_i w_i^\dagger + w_i \dot{w}_i^\dagger = 0.
\end{align}
The vertical space is then given by
\begin{equation}
    V_{w_i} = \{\dot{w}_i \in T_{w_i} V_{r_i}(\mathbb{C}^n) : \dot{w}_i w_i^\dagger + w_i \dot{w}_i^\dagger = 0 \}.
    \label{eq: Vert_condition}
\end{equation}
The projection $\pi_{i}$  has the derivative, $d\pi_i=w_idw_i^{\dagger}+dw_iw_i^{\dagger}$,  and the vertical tangent vectors are in the kernel of this linear map. Given a fiber of $\pi_i$ and a choice of a $w_i$ in this fiber, we can diffeomorphically identify the fiber with $\mbox{U}(r_i)$ by right multiplication. Suppose we take $X \in \mathfrak{u}(r_i)$, identified as an anti-Hermitian matrix in the usual way, and choose a curve $t\mapsto w_i(t) = w_i  \cdot e^{tX}$. Clearly, the projection onto the base is invariant under this transformation
\begin{align}
    w_i(t)w_i^\dagger(t) &= w_i \, e^{tX} \left(w_i \, e^{tX}\right)^\dagger \nonumber\\
    &= w_i \, e^{tX} e^{- t X} \,  w_i^\dagger = w_i w_i^\dagger.
\end{align}
The tangent vector to the fiber can now be written as $\dfrac{d w_i}{dt}\Big|_{t=0} = \dot{w}_i = w_i \cdot X$, which satisfies the condition for vertical matrices
\begin{align}
    \dot{w}_i w_i^\dagger + w_i \dot{w}_i^\dagger& = w_i X w_i^\dagger + w_i X^\dagger w_i \nonumber \\
    & = w_i X w_i^\dagger - w_i X w_i^\dagger \\
    &= 0. \nonumber
\end{align}
Hence, by dimensionality, our vertical space can also be seen as
\begin{equation}
   \!\!\!\! V_{w_i} \!\!=\! \{\dot{w}_i \!\in\! T_{w_i} V_{r_i}\!(\!\mathbb{C}^n\!) \!\!:\! \dot{w}_i \!=\! w_i \!\cdot\! X, \text{where} \ X^\dagger\!\! =\! - X \}.
\end{equation}
We are now in a condition to define the horizontal subspaces, which will simply be the collection of tangent vectors $\dot{w}_i$ that are orthogonal to $V_{w_i}$
\begin{align}\label{eq:H_space1}
    H_{w_i} &= (V_{w_i} )^{\perp}  \\
    &= \{ \dot{w} \in T_{w_i} V_{r_i}(\mathbb{C}^n) : \ \langle \dot{w}_i , \dot{w}_i^\prime \rangle = 0 , \text{where} \ \dot{w}_i^\prime \in V_{w_i} \}.\nonumber
\end{align}
Note that the operation $\langle \cdot , \cdot \rangle$ is not the Hermitian form defined in Eq.~\eqref{eq:hermitian_form}. It is instead the standard inner product in the space of complex matrices seen as a real vector space $\langle A , B \rangle \equiv \Re\Tr(A^\dagger B)$. The condition in \eqref{eq:H_space1} is then given by
\begin{align}
\Re\Tr \left( \dot{w}_i^\dagger w_i \cdot X \right)=0, & \text{ for every } X\in\mathfrak{u}(r_i) \nonumber \\
\implies &  \dot{w}_i^\dagger w_i - w_i^\dagger \dot{w}_i= 0 ,\label{eq: hor_condition}
\end{align}
where the implication stems from the fact that $X$ is anti-Hermitian, so that $\Dot{w}^\dagger w$ can only be Hermitian.~\footnote{To see this, observe that a complex matrix can be split into its Hermitian and anti-Hermitian components: $Z = Z^H + Z^{AH}$, where $Z^H = \frac{1}{2}(Z + Z^\dagger)$  and $Z^{AH} = \frac{1}{2}(Z - Z^\dagger)$. This real-linear decomposition divides the full matrix into two orthogonal components.  Indeed, $\Re\Tr\left[\left(Z_{1}^{A H}\right)^{\dagger} Z_{2}^{H}\right]=\ds{\frac{1}{2}}\left\{\Tr\left[\left(Z_{1}^{A H}\right)^{\dagger} Z_{2}^{H}\right]+\operatorname{Tr}\left[\left(Z_{2}^{H}\right)^{\dagger} Z_{1}^{A H}\right]\right\}=\ds{\frac{1}{2}}\left\{-\operatorname{Tr}\left[\left(Z_{1}^{A H}\right) Z_{2}^{H}\right]+\operatorname{Tr}\left[\left(Z_{2}^{H}\right) Z_{1}^{A H}\right]\right\}= 0$. Moreover, since the real vector space of Hermitian matrices and anti-Hermitian matrices both have dimension $k\times k$, we conclude that if a complex matrix is (real-)orthogonal to an anti-Hermitian matrix, then it must be Hermitian.} We can go further by making use of the condition in Eq.~\eqref{eq:T_space}, yielding $\dot{w}_i^\dagger w_i = - w_i^\dagger \dot{w}_i $; substituting this into Eq.~\eqref{eq: hor_condition} we get
\begin{equation}
    \dot{w}_i^\dagger w_i - w_i^\dagger \dot{w}_i = - 2 w_i^\dagger \dot{w}_i = 0  \implies w_i^\dagger \dot{w}_i = 0 .
\end{equation}

Finally, now that we have a notion of horizontal subspaces of the tangent spaces to $V_{r_i}(\mathbb{C}^n)$, we have unique isomorphisms of $ H_{w_i}\cong T_{P_i} \mbox{Gr}_{r_i}(\mathbb{C}^n) $ provided by the projection $\pi_i$. This means that for each $v  \in T_{P_i}\mbox{Gr}_{r_i}(\mathbb{C}^n)$  there exists a unique $\widetilde{v}^H\in H_{w_i}\subset T_{w_i} \mbox{V}_{r_i}(\mathbb{C}^n)$, such that its projection is $v$, i.e., $ \pi_i(\widetilde{v}^H) = \widetilde{v}^H w_i^\dagger +w_i\widetilde{v}^{H\dagger}=v$, and the converse is also true. This lift is called the ``horizontal lift'' for obvious reasons. Any other lift of $v$ to $T_{w_i}\mbox{V}_{r_i}(\mathbb{C}^n),$ i.e., any tangent vector projecting to $v$, would differ from the horizontal by an element of the kernel of the derivative of the projection, i.e., a vertical vector. As a consequence of this isomorphism, the Riemannian metric in the base space is  $g_i(v_1,v_2):=\langle \widetilde{v}_1^H, \widetilde{v}_2^H \rangle = \Re\Tr\left[\left(\widetilde{v}_1^H\right)^\dagger \widetilde{v}_2^H\right] $, where $\widetilde{v}_i^H$, are horizontal lifts of tangent vectors $v_1,v_2\in T_{P_i}\mbox{Gr}_{r_i}(\mathbb{C}^n)$. Moreover, the expression $g_i(v_1,v_2)$ does not depend on the point of the fiber over $P_i$, because the horizontal subspaces are $\mbox{U}(r_i)$-equivariant and the metric is $\mbox{U}(r_i)$-invariant. Indeed, if $\widetilde{v}^{H}\in H_{w_i}$ is a horizontal lift of $v\in T_{P_i}\mbox{Gr}_{r_i}(\mathbb{C}^n)$, then $\widetilde{v}^H\cdot U$ is a horizontal lift belonging to $H_{w_i\cdot U}$, for every $U\in \mbox{U}(r_i)$: $w_i^\dagger \widetilde{v}^{H}=0\Rightarrow (w_i\cdot U)^\dagger (\widetilde{v}^H\cdot U)= U^{\dagger}w_i^\dagger \widetilde{v}^H U=0$. Note that, in $\widetilde{v}^H\cdot U$, right multiplication should be understood as the tangent map of right multiplication at $w_i$. Finally, $\Re\Tr\left[\left(\widetilde{v}_1^H\right)^\dagger \widetilde{v}_2^H\right]=\Re\Tr\left[\left(\widetilde{v}_1^H\cdot U\right)^\dagger \widetilde{v}_2^H\cdot U\right]$, by the cyclic property of the trace, which shows that this expression defines a metric in the base space.

Now every tangent vector $\widetilde{v}\in T_{w_i}\mbox{V}_{r_i}(\mathbb{C}^n)$ is uniquely projected to a horizontal vector $\tilde{v}^H\in H_{w_i}$, which is mapped to a base space tangent vector $v\in T_{P_i}\mbox{Gr}_{r_i}(\mathbb{C}^n)$. Given the decomposition $T_{w_i} \mbox{V}_{r_i}(\mathbb{C}^n)=V_{w_i}\oplus H_{w_i}$, we can always find unique projection operators onto the vertical and horizontal subspaces, that perform the splitting 
\begin{align}
\widetilde{v}=\widetilde{v}^{V}+\widetilde{v}^{H}, \text{ where } \widetilde{v}^{V}\in V_{w_i}, \widetilde{v}^{H}\in H_{w_i}.
\end{align}
We have the identity
\begin{align}
g(v_1,v_2)=\langle \widetilde{v}_1^H, \widetilde{v}_2^H\rangle.
\end{align}
Additionally, due to the splitting of subspaces, we can write 
\begin{equation}
    \widetilde{v}^H = \widetilde{v} - \widetilde{v}^V.
\end{equation}

 In the following, we determine the form of the projection onto the vertical subspaces, in order to obtain a more compact form for the metric on the base space.

We claim that the vertical projection of a general tangent vector $\widetilde{v}$ is given by
\begin{equation}
 \widetilde{v}^V= P_i\widetilde{v} = w_i w_i^\dagger \widetilde{v}.
\end{equation}
Let us see why this is true. For this tangent vector to be vertical it must comply with Eq~\eqref{eq: Vert_condition}, i.e.,
\begin{equation}
    \left( P_i \widetilde{v}\right) w_i^\dagger + w_i  \left( P_i \widetilde{v} \right)^\dagger = w_i w_i^\dagger \widetilde{v} w_i^\dagger + w_i \widetilde{v}^\dagger w_i w_i^\dagger = 0.
\end{equation}
However, we know that $\widetilde{v}$ is a tangent vector, that is, we know that $\widetilde{v}^\dagger w_i = - w_i^\dagger \widetilde{v}$. Replacing this in the expression above we have
\begin{equation}
    w_i w_i^\dagger \widetilde{v} w_i^\dagger - w_i w_i^\dagger \widetilde{v} w_i^\dagger = 0.
\end{equation}
Hence, we have verified that $P_i \widetilde{v}$ is a vertical tangent vector and the map $\widetilde{v} \mapsto w_i w_i^\dagger \widetilde{v}$ is a projection onto the vertical space. The horizontal projection is then given by
\begin{equation}
    \widetilde{v}^H = \widetilde{v} - (w_i w_i^\dagger) \widetilde{v}.
\end{equation}
Meanwhile, the metric in $\mbox{Gr}_{r_i}(\mathbb{C}^n)$ is, using the horizontal projections, given by the following compact formula
\begin{align}
    g_i & = \Re\Tr\left[ \left(dw_i^\dagger - dw_i^\dagger w_iw_i^\dagger\right) \left( dw_i - w_iw_i^\dagger dw_i\right) \right] \nonumber\\
& = \Re\Tr\Big[ dw_i^\dagger dw_i -  dw_i^\dagger w_iw_i^\dagger dw_i\nonumber \\
&- dw_i^\dagger w_iw_i^\dagger dw_i + dw_i^\dagger (w_iw_i^\dagger)^2 dw_i  \Big]. 
\end{align}
We know that $ w_i(w_i^\dagger w_i ) w_i^\dagger = w_iw_i^\dagger$, since $w_i^\dagger w_i = I_k$, so the last two terms cancel each other, giving
\begin{align}
    g_i &= \Re\Tr\left[ dw_i^\dagger dw_i -  dw_i^\dagger w_iw_i^\dagger dw_i  \right] \nonumber\\
    &= \Re\Tr\left[ dw_i^\dagger( 1 - w_iw_i^\dagger) dw_i \right]. \label{eq:Metric1}
\end{align}

Now, this expression is written in terms of the elements defined in the principal bundle so we want to write it in terms of the elements in the base space --- the projectors~$P_i$. For this purpose, notice that $w_i = (w_i w_i^\dagger) w_i = P_i w_i$ which, by derivation gives $d w_i = dP_i w_i + P_i dw_i$. The same can be done for the hermitian $w_i^\dagger = w_i^\dagger (w_i w_i^\dagger) = w_i^\dagger P_i $ which gives us $dw_i^\dagger = dw_i^\dagger P_i + w_i^\dagger d P_i$. Replacing these in Eq.~\eqref{eq:Metric1}, we get
\begin{align}
    g_i  = \Re \Tr & \left[dw_i^\dagger ( 1 - w_i w_i^\dagger) dw_i \right] \nonumber \\
     = \Re \Tr & \left[\left(dw_i^\dagger P_i + w_i^\dagger dP_i\right) \left( 1 - P_i \right) \left (  dP_i w_i + P_idw_i \right)\right] \nonumber \\
     = \Re \Tr & \Big[ \left ( dw_i^\dagger P_i + w_i^\dagger dP_i - dw_i^\dagger P_i - w_i^\dagger dP_i P_i \right) \nonumber\\
     & \cdot \left( dP_iw_i + P_i dw_i \right) \Big] \nonumber \\
     = \Re \Tr & \Big( dw_i^\dagger P_i dP_i w_i + w_i^\dagger dP_i dP_i w_i  \\
     &- dw_i^\dagger P_i dP_i w_i - w_i^\dagger dP_i P_i dP_i w_i \nonumber \\
    & + dw_i^\dagger P_i  dw_i + w_i^\dagger dP_i P_i dw_i \nonumber\\
    &- dw_i^\dagger P_i  dw_i - w_i^\dagger dP_i P_i dw_i \Big) \nonumber \\
     = \Re \Tr & \Big( w_i^\dagger dP_i dP_i w_i  - w_i^\dagger dP_i P_i dP_i w_i \Big)  \nonumber \\
     = \Re \Tr & \Big( P_i dP_i dP_i \Big)  - \Re \Tr\Big( P_i dP_i P_i dP_i\Big). \nonumber
\end{align}
Moreover, since $P_i^2 = P_i$, we have that $dP_i = d(P_i^2) = P_i dP_i + dP_i P_i $. Multiplying this expression by $P_i$ on both sides we get $P_i dP_i P_i = 2 P_i dP_i P_i$ and we can conclude that $P_i dP_i P_i = 0 $. The last term on the last expression is then zero and we see that the metric is given by 
\begin{align}\label{eq:Grassmannian_metric}
    g_i &= \Re \Tr\left( P_i dP_i dP_i \right)=\Re \Tr\left( P_i dP_i dP_i P_i \right)\nonumber \\
    &=\Tr\left( P_i dP_i dP_i P_i \right).
\end{align}

Now we wish to determine the metric on the total space of the principal bundle, i.e., the metric that encompasses both the classical and quantum parts. For this purpose, consider a curve in the principal bundle space given by $t \mapsto p_\tau(t) = \left( \sqrt{\bf{p}(t)}, \bf{w}(t)\right)$ and compute the distance between two infinitesimally close points $t$ and $t + \delta t$. For the first case, we consider a static $\bf{w}(t) = \bf{w}$ and compute the distance

\begin{align}
d^2_{\tau}\big(p_\tau(t), & \ p_\tau(t + \delta t)\big) \\ 
& = 2 \Big( 1 - \sum_{i=1}^k \sqrt{p_i(t)p_i(t + \delta t)} \Re\Tr(w_i^\dagger w_i) \Big).\nonumber
\end{align}
We have $\Tr(w_i^\dagger w_i)=\Tr P_i = r_i$, hence
\begin{align}
\!\!\!\!d^2_{\tau}\big(p_\tau(t), p_\tau(t \!+\! \delta t)\big) \!=\! 2 \!\left(\!\! 1 \!-\! \sum_{i=1}^k r_i \sqrt{p_i(t)p_i(t \!+\! \delta t)}\! \right)\!.\label{eq:inf_distance}
\end{align}
Let us look more closely at the expression $ \sqrt{p_i(t)p_i(t + \delta t)}$. We can Taylor expand $P_i(t+\delta t)$ to second order in $\delta t$ to obtain
\begin{equation}
\begin{array}{rcl}
\!\!\!\!\!\!\!\!\sqrt{p_i(t)p_i(t + \delta t)} \!\!\! & = & \!\!\! \ds{\sqrt{p_i(t)\left(p_i(t) + \dot{p}_i \delta t + \frac{1}{2} \ddot{p}_i \delta t^2\right)}}\\[0.2cm]
\!\!\! & = & \!\!\! \ds{p_i(t) \sqrt{1 + \frac{\dot{p}_i}{p_i} \delta t + \frac{1}{2}\frac{\ddot{p}_i}{p_i} \delta t^2}}
\end{array}
\end{equation}
We can then approximate the quantity inside the square root by $\sqrt{1 + x} \approx 1 + \frac{1}{2} x - \frac{1}{8} x^ 2$, which, ignoring higher order terms, yields
\begin{align}
&\sqrt{p_i(t)p_i(t + \delta t)} \nonumber \\
&\approx  p_i\Big[1 + \frac{1}{2} \left( \frac{\dot{p}_i}{p_i} \delta t + \frac{1}{2}\frac{\ddot{p}_i}{p_i} \delta t^2 \right)- \frac{1}{8} \left( \frac{\dot{p}_i}{p_i} \delta t + \frac{1}{2}\frac{\ddot{p}_i}{p_i} \delta t^2 \right)^2 \Big]\nonumber\\
& =   p_i\left[1  +\frac{1}{2}  \frac{\dot{p}_i}{p_i} \delta t + \frac{1}{2}  \frac{\ddot{p}_i}{p_i} \delta t^2  -  \frac{1}{8}   \left(\frac{\dot{p}_i}{p_i}\right)^2 \delta t^2 \right]\\
&=  p_i +\frac{1}{2} \dot{p}_i \delta t + \frac{1}{2} \ddot{p}_i \delta t^2  -  \frac{1}{8} \frac{\dot{p}_i^2}{p_i}\delta t^2 . \nonumber
\end{align}
Replacing this in Eq.~\eqref{eq:inf_distance}, we get 
\begin{align}
&d^2_{\tau}\left(p_\tau(t), p_\tau(t + \delta t)\right) \\
& = 2 \Big[ 1 -\sum_{i=1}^k r_i \left(p_i +\frac{1}{2} \dot{p}_i \delta t + \frac{1}{2} \ddot{p}_i \delta t^2  -  \frac{1}{8} \frac{\dot{p}_i^2}{p_i}\delta t^2 \right) \Big]. \nonumber
\end{align}
Using the condition $ \sum_{i=1}^k r_i  p_i = 1$ we can infer that $\sum_{i=1}^k r_i \dot{p}_i = 0$ and $\sum_{i=1}^k r_i \ddot{p_i} = 0 $. Applying these results in the expression above, we finally arrive at the Fisher-Rao metric 
\begin{align}
\left(ds_{P}^{\text{Cl}}\right)^2 & = \frac{1}{4} \sum_{i=1}^k r_i  \frac{\dot{p}_i^2}{p_i}\delta t^2 = \frac{1}{4} \sum_{i=1}^k r_i  \frac{dp_i^2}{p_i}
\end{align}
in terms of the probability distribution ``coordinates''~$\sqrt{\bf{p}}$.

Next, consider the case of a static classical part~$\bf{p}(t) = \bf{p}$. The distance is then 
\begin{align}
\label{eq:inf_distance2}
d^2_{\tau}\Big(p_\tau(t), & \ p_\tau(t + \delta t)\Big) \\
& = 2 \Big( 1 - \sum_{i=1}^k p_i \Re\Tr(w_i(t)^\dagger w_i(t+ \delta t)) \Big). \nonumber
\end{align}
Expanding $w_i(t+\delta t)$ to second order $w_i(t + \delta t) \approx w_i(t) + \dot{w}_i(t) \delta t + \frac{1}{2} \ddot{w}_i(t) \delta t^2 $ we have
\begin{align}\label{eq:Tay_expansion}
&\Re\Tr(w_i(t)^\dagger w_i(t+ \delta t))\\
&= \Re \Tr(w_i^\dagger w_i)  + \Re \Tr (w_i^\dagger \dot{w}_i) \delta t + \frac{1}{2} \Re \Tr (w_i^\dagger \ddot{w}_i) \delta t^2\nonumber \\
 &+ r_i  + \frac{1}{2} \Tr (w_i^\dagger \dot{w}_i + \dot{w}_i^\dagger w_i ) \delta t + \frac{1}{4}  \Tr (w_i^\dagger \ddot{w}_i + \ddot{w}_i^\dagger w_i  ) \delta t^2.	 \nonumber
\end{align}
From condition \eqref{eq:T_space} for tangent vectors, the first order term is zero. From this same condition one can infer that $ \ddot{w}_i^\dagger w_i + w_i^\dagger \ddot{w}_i = -2 \dot{w}_i^\dagger \dot{w}_i  $ and Eq.~\eqref{eq:Tay_expansion} becomes
\begin{align}
\Re\Tr\big(w_i(t)^\dagger w_i(t+ \delta t)\big) \! =  r_i  - \frac{1}{2} \Tr\left(\dot{w}_i^\dagger \dot{w}_i\right) \delta t^2.
\end{align}
Using this expression in Eq.~\eqref{eq:inf_distance2} we get
\begin{align}
d^2_{\tau}\big(p_\tau(t), & \ p_\tau(t + \delta t)\big) \\
& = 2 \Big( 1
-\sum_{i=1}^k r_i p_i + \frac{1}{2} \sum_{i=1}^k  p_i \Tr(\dot{w}_i^\dagger \dot{w}_i) \delta t^2\Big). \nonumber
\end{align}
Since $\sum_{i=1}^k r_i p_i = 1 $, we have
\begin{equation}
\!\!\!\!\!\left(ds_{P_{\tau}}^{\text{Q}}\right)^{2} \!=\! \sum_{i=1}^k p_i \Tr(\dot{w}_i^\dagger \dot{w}_i) \delta t^2  \!=\!    \sum_{i=1}^k p_i \Tr(dw_i^\dagger dw_i).
\end{equation}
From the derivation of Eq.~\eqref{eq:Grassmannian_metric}, it becomes clear that, restricting to the horizontal subspaces, one obtains the induced quantum part of the metric in the base space
\begin{equation}
\left(ds_{B_{\tau}}^{\text{Q}}\right)^{2} =  \sum_{i=1}^k p_i \Tr\left( P_i dP_i dP_i \right).
\end{equation}
So, the quantum part of the metric in the base space is the sum for $i\in\{1,...,k\}$ of the metric on the Grassmannian given by Eq.~\eqref{eq:Grassmannian_metric} weighed by the relative proportions of the distribution $p_i$.

\medskip

Finally, we are left with the task of taking a general variation, where both $\sqrt{\bf {p}}(t)$ and $\bf{w}(t)$ are non-constant, to make sure that we do not get cross terms. We have,
\begin{align}
d^2_{\tau} & \big(p_\tau(t), p_\tau(t + \delta t)\big) \\
& = 2 \Big( 1 - \sum_{i=1}^k \sqrt{p_i(t)p_i(t + \delta t)} \Re\Tr(w_i(t)^\dagger w_i(t+ \delta t))  \Big). \nonumber
\end{align}
We can Taylor expand, as before, to obtain
\begin{align}
d^2_{\tau}  \big(p_\tau(t), & \ p_\tau(t + \delta t)\big)  \\
= 2 \Big[ 1 & - \sum_{i=1}^k \Big(p_i +\frac{1}{2} \dot{p}_i \delta t + \frac{1}{2} \ddot{p}_i \delta t^2 \nonumber \\
&-  \frac{1}{8} \frac{\dot{p}_i^2}{p_i}\delta t^2  \big( r_i - \frac{1}{2}\Tr(\dot{w}_i^\dagger \dot{w}_i) \delta t^2 \big)\Big)  \Big]. \nonumber
\end{align}
Collecting the terms up to second order, we get
\begin{align}
d^2_{\tau}\big(p_\tau(t), p_\tau(t + \delta t)\big) & \\
= 2 \Big[ 1 - \sum_{i=1}^k \Big( & p_i \Tr(\dot{w}_i^\dagger \dot{w}_i) \delta t^2 + \frac{1}{2} r_i \dot{p}_i \delta t \nonumber \\ 
& + \frac{1}{2} r_i \ddot{p}_i \delta t^2 - \frac{1}{8} r_i \frac{\dot{p}_i^2}{p_i}\delta t^2 \Big) \Big], \nonumber
\end{align}
which, using the same arguments as before, reduces to
\begin{align}
ds_{P_{\tau}}^{2} &=  \sum_{i=1}^k \left(   \frac{1}{4} r_i \frac{\dot{p}_i^2}{p_i}\delta t^2  +  p_i \Tr(\dot{w}_i^\dagger \dot{w}_i) \delta t^2  \right) \nonumber\\
&=  \sum_{i=1}^k \left(  \frac{1}{4} r_i \frac{dp_i^2}{p_i} + p_i \Tr(dw_i^\dagger dw_i)   \right).
\end{align}
Hence, the metric in the principal bundle is just the sum of the respective classical and quantum metrics. We want to arrive at the metric for the base space: The classical probability distributions $\sqrt{p_i}$ have no gauge freedom so they have no vertical or horizontal components and their projection is trivial; meanwhile, the horizontal projection in the quantum part described by the amplitudes $w_i$ proceeds as in the Stiefel manifold case, for each $i=1,...,k$, so that our final interferometric metric $g_{I}$ is
\begin{align}
g_{I}&=ds_{B_{\tau}}^{2} \nonumber \\
&= \left(ds_{B_\tau}^{\text{Cl}}\right)^{2} + \left(ds_{B_\tau}^{\text{Q}}\right)^{2} \\
&=    \frac{1}{4} \sum_{i=1}^k   r_i   \frac{dp_i^2}{p_i} +\sum_{i=1}^k  \ p_i  \Tr\left( P_i dP_i dP_i \right)\nonumber  .
\end{align}
\section{The proof of the maximal output probability in the interferometric experiment}
\label{sec:interferometric_probability}
The input state is $\ket 0 \bra 0 \otimes \rho$. The first beam splitter $\mbox{BS1} \otimes I$ acts on this state giving $\frac 1 2\left( \ket 0 + i \ket 1 \right) \left( \bra 0 - i \bra 1 \right)\otimes \rho$. The controlled unitary is $ \ket 0 \bra 0 \otimes V + \ket 1 \bra 1 \otimes U$, which, when acting on the last state, gives 
\begin{equation}
\begin{array}{rl}
\displaystyle \frac{1}{2} \Big( & \!\!\! \ket 0 \bra 0 \otimes V \rho V^\dagger - i  \ket 0 \bra 1 \otimes V \rho U^\dagger  \\ 
  & + i \ket 1 \bra 0 \otimes U \rho V^\dagger + \ket 1 \bra 1 \otimes U \rho U^\dagger \Big).
\end{array}
\end{equation}
Passing through a second beam splitter and measuring the $\ket 1$ state yield
\begin{align}
\frac 1 4 \ket 1 \bra 1 \otimes \Big[& V \rho V^\dagger  + V \rho U^\dagger    + U \rho V^\dagger +   U \rho U^\dagger \Big].
\end{align}
Tracing out this quantity gives
\begin{align}
\frac 1 4 \left[ \Tr U \rho U^\dagger + \Tr V\rho V^\dagger + 2 \Re\Tr U\rho V^\dagger \right].
\end{align}
We know that $\Tr U \rho U^\dagger = \Tr V\rho V^\dagger = 1$, hence
\begin{align}
&\frac 1 2 \left[ 1 +  \Re\Tr U\rho V^\dagger \right] .
\end{align}
Recall that $V= \sum_{i=0}^k P_i VP_i$, and that since we can write, in terms of a choice of amplitudes $w_i$, $i=1,...,k$,
\begin{align}
P_i=w_iw_i^{\dagger}, \ i=1,...,k,
\end{align}
then,
\begin{align}
\label{eq: V}
V=P_0VP_0 +\sum_{i=1}^{k}w_i V_iw_i^\dagger,
\end{align}
where $V_i=w_{i}^{\dagger} V w_i$ is an $r_i\times r_i$ unitary matrix, for $i=1,...,k$. Observe that 
\begin{align}
\tr \big[V^{\dagger}U \rho \big]&=\sum_{i,j=0}^{k}p_i \tr \big[P_jV^{\dagger} P_j U P_i\big] \nonumber\\
&=\sum_{i=0}^{k}p_i \tr \big[V^{\dagger} P_i U P_i\big], 
\end{align}
where in the last step we used the cyclic property of the trace and $P_iP_j=\delta_{ij}P_i$, $i,j=0,...,k$. Finally, introducing the expression for $V$ of Eq.~\eqref{eq: V} we can write, using $w_i^\dagger w_i=I_{r_i}$, $i=1,..,r_i$, and $p_0=0$,
\begin{align}
&\sum_{i=1}^{k}p_i \tr \big[V^{\dagger} P_i U P_i\big]\nonumber\\
&=\sum_{i=1}^{k}p_i \tr \big[(V_i^{\dagger} w_i^\dagger U) w_i\big]\\
&=\sum_{i=1}^{k}p_i \tr \big[(U^{\dagger} w_iV_i)^{\dagger} w_i\big]. \nonumber
\end{align}
Observe that if we write
\begin{align}
p_{\tau}=((p_i,w_i))_{i=1}^{k} \text{ and } q_{\tau}=((p_i, U^{\dagger} w_iV_i))_{i=1}^{k},
\end{align}
then, 
\begin{align}
\sum_{i=1}^{k}p_i \tr \big[V_i^{\dagger} w_i^\dagger U w_i\big]=\langle q_{\tau},p_{\tau}\rangle_{\tau},
\end{align}
where $\langle q_{\tau},p_{\tau}\rangle$ is the Hermitian form defined in Eq.~\eqref{eq:hermitian_form}. Hence,
\begin{align}
\pr_1 &=\frac{1}{2}\left(1+\Re \tr U\rho V^\dagger \right)\nonumber\\
&= 1-\frac{1}{2}\left(1-\sum_{i=1}^{k}p_i\Re \tr \big[P_iV^{\dagger} P_i U P_i\big]\right)\nonumber\\
&= 1-\frac{1}{2}\left(1-\sum_{i=1}^{k}p_i\Re \langle q_{\tau},p_{\tau}\rangle_{\tau} \right)\\
&=1- \frac{1}{4}d^{2}_{\tau}(q_{\tau},p_{\tau}),\nonumber
\end{align}
where $d_{\tau}$ is the distance over the total space of the principal bundle $P_{\tau}\to B_{\tau}$. Maximizing over the the gauge degree of freedom given by the collection of unitary $r_i\times r_i$ matrices, $V_i$, $i=1,...,k$ (note that $P_0VP_0$ is irrelevant), one gets the distance $d_{I}(\rho,U^\dagger\rho U)$ over the base space $B_{\tau}$ of as explored in the main text.
\section{Pullback of interferometric metric to parameter space}
\label{sec: pullback of interferometric metric to parameter space}
We wish to find the metric obtained by pulling back the interferometric metric
\begin{align}
&g=\frac{1}{4}\sum_i r_i \frac{dp_i^2}{p_i}+\sum_{i}p_i \Tr\left(P_idP_idP_i\right), \nonumber\\
&\text{with }  \rho=\sum_i p_i P_i \text{ and} \Tr P_i=r_i,
\end{align}
by the map induced by the Gibbs state
\begin{align}
M \mapsto \rho(M)=Z^{-1}\exp(-\beta \mathcal{H}(M)),
\end{align}
with $\mathcal{H}(M)$ given by Eq.~\eqref{eq:band insulator} and where $Z$ is the partition function. The first thing to notice is that if $\rho=\rho_1\otimes \rho_2$, with $\rho_\alpha=\sum_{i_{\alpha}}p_{i_{\alpha}}P_{i_{\alpha}}$, $\alpha=1,2$ we have the decomposition
\begin{align}
\rho=\sum_{I}p_{I}P_{I}=\sum_{i_{1},i_{2}}p_{i_1}p_{i_2}P_{i_1}\otimes P_{i_2},
\end{align}
where $I=(i_1,i_2)$ is a multi-index describing the joint system labels. Note that,
\begin{align}
&\sum_{i_{1},i_{2}}r_{I}\frac{dp_{I}^2}{p_I}\nonumber\\
&=\sum_{i_{1},i_{2}}\frac{r_{i_{1}}r_{i_{2}}}{p_{i_{1}}p_{i_{2}}}\left(p_{i_{2}}^2 dp_{i_{1}}dp_{i_{1}}+2 p_{i_{1}} p_{i_{2}} dp_{i_{1}}dp_{i_{2}} +p_{i_{1}}^2 dp_{i_2}^2\right)\nonumber\\
&=\sum_{i_{1}}r_{i_{1}}\frac{dp^2_{i_{1}}}{p_{i_{1}}}+\sum_{i_{2}}r_{i_{2}}\frac{dp^2_{i_{2}}}{p_{i_{2}}},
\end{align}
and
\begin{align}
&\sum_{I}p_I \Tr\left(P_IdP_IdP_I\right)\\
&=\sum_{i_{1},i_{2}}p_{i_{1}} p_{i_{2}}\Tr\left[P_{i_1}\otimes P_{i_2}d\left(P_{i_1}\otimes P_{i_2}\right)d\left(P_{i_1}\otimes P_{i_2}\right)\right]\nonumber\\
&=\sum_{i_{1}}p_{i_1}\Tr\left(P_{i_1}dP_{i_1}dP_{i_1}\right)+\sum_{i_{2}}p_{i_2}\Tr\left(P_{i_2}dP_{i_2}dP_{i_2}\right), \nonumber
\end{align}
where we used $PdPP=0$ for any projector $P$. As a consequence, the interferometric metric, much like the Bures metric, converts tensor product states into orthogonal sum metrics.

Because the Hamiltonian is diagonal in momentum space, the density matrix factors over the momenta: It follows that the metric becomes an integral over the momentum space of individual contributions of each momentum sector. The pullback of the classical term, which also appears in the Bures metric,
\begin{align}
\frac{1}{4}\sum_{i}r_i\frac{dp_i^2}{p_i}
\end{align}
was computed in the Appendix of Ref.~\cite{ami:mer:vla:pau:vie:18} and it yields
\begin{align}
\frac{\beta^2}{4}\int_{\BZ^d}\frac{d^dk}{(2\pi)^d}\; \frac{1}{\cosh(\beta E(\bf{k};M))+1}\left(\frac{\partial E(\bf{k};M)}{\partial M}\right)^2  dM^2,
\end{align}
where $E(\bf{k};M)=|d(\bf{k};M)|$ is the magnitude of $d(\bf{k},M)$. With regard to the second term, one can use the mathematical fact that the embedding of the space of $k$-dimensional subspaces of $\mathbb{C}^N$, $\mbox{Gr}_k(\mathbb{C}^N)$ on the space of one-dimensional subspaces of the Fock space $\mathbb{P}\Lambda^*\mathbb{C}^N$, given by 
\begin{align}
\mbox{span}\left\{\ket{1},...,\ket{k}\right\}\mapsto \mbox{span}\left\{ c_{1}^\dagger ... c_{k}^\dagger \ket{0}\right\},
\end{align}
is isometric. In the previous equation $c_{i}^\dagger$ stand for creation operators for $\ket{i}$,that is, at the single particle level, $c_{i}^\dagger\ket{0}=\ket{i}$, $i=1,...,k$. The embedding being isometric means, in this context, that if we write the rank $k$ single-particle projector 
\begin{align}
\widetilde{P}=\sum_{i=1}^{k}\ket{i}\bra{i}
\end{align}
and the rank $1$ many-body projector
\begin{align}
P=c_{1}^\dagger ... c_{k}^\dagger \ket{0}\bra{0}c_{k}...c_{1},
\end{align}
we have
\begin{align}
\Tr\left(\widetilde{P} d\widetilde{P} d\widetilde{P}\right)= \Tr\left(P dP dP\right).
\end{align}
In particular, this means that in the gapped case for each $\bf{k}\in\BZ^d$ we will have four classes of orthogonal eigenstates,
\begin{align}
\ket{0},\; c^{\dagger}_{1,\bf{k}}\ket{0}, \; c^{\dagger}_{2,\bf{k}}\ket{0}, c^{\dagger}_{1,\bf{k}}c^{\dagger}_{2,\bf{k}}\ket{0},
\end{align}
where $c^{\dagger}_{i,\bf{k}}$, $i=1,2$, are the Bogoliubov quasiparticle creation operators of $\mathcal{H}$ with energies $E(\bf{k};M)$ and $-E(\bf{k};M)$, respectively. The energies of the classes of eigenstates are $0$, $E(\bf{k};M)$, $-E(\bf{k};M)$ and $0$, respectively. The associated single-particle $2\times 2$ projectors are the $0$ projector, $P_1(\bf{k};M)=c^{\dagger}_{1,\bf{k}}\ket{0}\bra{0}c_{1,\bf{k}}$, $P_{2}(\bf{k};M)=c^{\dagger}_{2,\bf{k}}\ket{0}\bra{0}c_{2,\bf{k}}$ and the $2\times 2$ identity matrix $I_2$, respectively. Only $P_1(\bf{k})$ and $P_2(\bf{k})$ are non-trivial (not constant), and moreover, if we introduce the unit vector $n=d/|d|$, we can write
\begin{align}
P_{1}(\bf{k};M)&=\frac{1}{2}\left(I_2+n^\mu(\bf{k};M)\sigma_\mu\right)\nonumber \\
P_2(\bf{k};M)&=I_2-P_{1}(\bf{k};M).
\end{align}
As a consequence, using the identity $\Tr\left(PdPdP\right)=(1/2)\Tr\left(dPdP\right)$ and using the fact that the Pauli matrices are traceless, we get, 
\begin{align}
\Tr \left(P_1dP_1dP_1\right)&=\Tr \left(P_2dP_2dP_2\right)\\
&=\frac{1}{4}\delta_{\mu\nu}\frac{\partial n^\mu(\bf{k};M)}{\partial M}\frac{\partial n^\nu(\bf{k};M)}{\partial M}dM^2.\nonumber
\end{align}
Finally, taking into account the partition function factor $Z_{\bf{k}}=\left(2+2\cosh(\beta E(\bf{k};M))\right)$, we get that the quantum contribution is
\begin{align*}
\frac{1}{4}\int_{\BZ^d}\frac{d^d k}{(2\pi)^d}&\left( \frac{\cosh(\beta E(\bf{k};M))}{1+\cosh(\beta E(\bf{k};M))}\right)\nonumber\\
&\times \delta_{\mu\nu}\frac{\partial n^\mu(\bf{k};M)}{\partial M}\frac{\partial n^\nu(\bf{k};M)}{\partial M}dM^2.
\end{align*}
Finally, we obtain
\begin{align}
g&=\frac{1}{4}\int_{\BZ^d}\frac{d^d k}{(2\pi)^d}\Big[ \frac{1}{\cosh(\beta E)+1}\\
&\times \left(\beta^2\left(\frac{\partial E}{\partial M}\right)^2 + \cosh(\beta E)\delta_{\mu\nu}\frac{\partial n^\mu}{\partial M}\frac{\partial n^\nu}{\partial M}\right)\Big]dM^2,\nonumber
\end{align}
where we omitted the obvious dependence on $\bf{k}$ and $M$ of the quantities $E$ and $n^\mu$.

\bibliography{bib}

\begin{thebibliography}{36}%
\makeatletter
\providecommand \@ifxundefined [1]{%
 \@ifx{#1\undefined}
}%
\providecommand \@ifnum [1]{%
 \ifnum #1\expandafter \@firstoftwo
 \else \expandafter \@secondoftwo
 \fi
}%
\providecommand \@ifx [1]{%
 \ifx #1\expandafter \@firstoftwo
 \else \expandafter \@secondoftwo
 \fi
}%
\providecommand \natexlab [1]{#1}%
\providecommand \enquote  [1]{``#1''}%
\providecommand \bibnamefont  [1]{#1}%
\providecommand \bibfnamefont [1]{#1}%
\providecommand \citenamefont [1]{#1}%
\providecommand \href@noop [0]{\@secondoftwo}%
\providecommand \href [0]{\begingroup \@sanitize@url \@href}%
\providecommand \@href[1]{\@@startlink{#1}\@@href}%
\providecommand \@@href[1]{\endgroup#1\@@endlink}%
\providecommand \@sanitize@url [0]{\catcode `\\12\catcode `\$12\catcode
  `\&12\catcode `\#12\catcode `\^12\catcode `\_12\catcode `\%12\relax}%
\providecommand \@@startlink[1]{}%
\providecommand \@@endlink[0]{}%
\providecommand \url  [0]{\begingroup\@sanitize@url \@url }%
\providecommand \@url [1]{\endgroup\@href {#1}{\urlprefix }}%
\providecommand \urlprefix  [0]{URL }%
\providecommand \Eprint [0]{\href }%
\providecommand \doibase [0]{https://doi.org/}%
\providecommand \selectlanguage [0]{\@gobble}%
\providecommand \bibinfo  [0]{\@secondoftwo}%
\providecommand \bibfield  [0]{\@secondoftwo}%
\providecommand \translation [1]{[#1]}%
\providecommand \BibitemOpen [0]{}%
\providecommand \bibitemStop [0]{}%
\providecommand \bibitemNoStop [0]{.\EOS\space}%
\providecommand \EOS [0]{\spacefactor3000\relax}%
\providecommand \BibitemShut  [1]{\csname bibitem#1\endcsname}%
\let\auto@bib@innerbib\@empty
\bibitem [{\citenamefont {Haldane}(1988)}]{hal:88}%
  \BibitemOpen
  \bibfield  {author} {\bibinfo {author} {\bibfnamefont {F.~D.~M.}\
  \bibnamefont {Haldane}},\ }\bibfield  {title} {\bibinfo {title} {Model for a
  quantum hall effect without landau levels: Condensed-matter realization of
  the" parity anomaly"},\ }\href@noop {} {\bibfield  {journal} {\bibinfo
  {journal} {Physical review letters}\ }\textbf {\bibinfo {volume} {61}},\
  \bibinfo {pages} {2015} (\bibinfo {year} {1988})}\BibitemShut {NoStop}%
\bibitem [{\citenamefont {Kitaev}(2009)}]{kit:09}%
  \BibitemOpen
  \bibfield  {author} {\bibinfo {author} {\bibfnamefont {A.}~\bibnamefont
  {Kitaev}},\ }\bibfield  {title} {\bibinfo {title} {Periodic table for
  topological insulators and superconductors},\ }in\ \href@noop {} {\emph
  {\bibinfo {booktitle} {AIP conference proceedings}}},\ Vol.\ \bibinfo
  {volume} {1134}\ (\bibinfo {organization} {American Institute of Physics},\
  \bibinfo {year} {2009})\ pp.\ \bibinfo {pages} {22--30}\BibitemShut {NoStop}%
\bibitem [{\citenamefont {Shiozaki}\ and\ \citenamefont
  {Sato}(2014)}]{shi:sat:14}%
  \BibitemOpen
  \bibfield  {author} {\bibinfo {author} {\bibfnamefont {K.}~\bibnamefont
  {Shiozaki}}\ and\ \bibinfo {author} {\bibfnamefont {M.}~\bibnamefont
  {Sato}},\ }\bibfield  {title} {\bibinfo {title} {Topology of crystalline
  insulators and superconductors},\ }\href
  {https://doi.org/10.1103/PhysRevB.90.165114} {\bibfield  {journal} {\bibinfo
  {journal} {Phys. Rev. B}\ }\textbf {\bibinfo {volume} {90}},\ \bibinfo
  {pages} {165114} (\bibinfo {year} {2014})}\BibitemShut {NoStop}%
\bibitem [{\citenamefont {Gong}\ \emph {et~al.}(2018)\citenamefont {Gong},
  \citenamefont {Ashida}, \citenamefont {Kawabata}, \citenamefont {Takasan},
  \citenamefont {Higashikawa},\ and\ \citenamefont {Ueda}}]{gon:ash:18}%
  \BibitemOpen
  \bibfield  {author} {\bibinfo {author} {\bibfnamefont {Z.}~\bibnamefont
  {Gong}}, \bibinfo {author} {\bibfnamefont {Y.}~\bibnamefont {Ashida}},
  \bibinfo {author} {\bibfnamefont {K.}~\bibnamefont {Kawabata}}, \bibinfo
  {author} {\bibfnamefont {K.}~\bibnamefont {Takasan}}, \bibinfo {author}
  {\bibfnamefont {S.}~\bibnamefont {Higashikawa}},\ and\ \bibinfo {author}
  {\bibfnamefont {M.}~\bibnamefont {Ueda}},\ }\bibfield  {title} {\bibinfo
  {title} {Topological phases of non-hermitian systems},\ }\href@noop {}
  {\bibfield  {journal} {\bibinfo  {journal} {Physical Review X}\ }\textbf
  {\bibinfo {volume} {8}},\ \bibinfo {pages} {031079} (\bibinfo {year}
  {2018})}\BibitemShut {NoStop}%
\bibitem [{\citenamefont {Roy}\ and\ \citenamefont
  {Harper}(2017)}]{roy:har:17}%
  \BibitemOpen
  \bibfield  {author} {\bibinfo {author} {\bibfnamefont {R.}~\bibnamefont
  {Roy}}\ and\ \bibinfo {author} {\bibfnamefont {F.}~\bibnamefont {Harper}},\
  }\bibfield  {title} {\bibinfo {title} {Periodic table for {F}loquet
  topological insulators},\ }\href@noop {} {\bibfield  {journal} {\bibinfo
  {journal} {Physical Review B}\ }\textbf {\bibinfo {volume} {96}},\ \bibinfo
  {pages} {155118} (\bibinfo {year} {2017})}\BibitemShut {NoStop}%
\bibitem [{\citenamefont {Schindler}\ \emph {et~al.}(2018)\citenamefont
  {Schindler}, \citenamefont {Cook}, \citenamefont {Vergniory}, \citenamefont
  {Wang}, \citenamefont {Parkin}, \citenamefont {Bernevig},\ and\ \citenamefont
  {Neupert}}]{sch:18}%
  \BibitemOpen
  \bibfield  {author} {\bibinfo {author} {\bibfnamefont {F.}~\bibnamefont
  {Schindler}}, \bibinfo {author} {\bibfnamefont {A.~M.}\ \bibnamefont {Cook}},
  \bibinfo {author} {\bibfnamefont {M.~G.}\ \bibnamefont {Vergniory}}, \bibinfo
  {author} {\bibfnamefont {Z.}~\bibnamefont {Wang}}, \bibinfo {author}
  {\bibfnamefont {S.~S.~P.}\ \bibnamefont {Parkin}}, \bibinfo {author}
  {\bibfnamefont {B.~A.}\ \bibnamefont {Bernevig}},\ and\ \bibinfo {author}
  {\bibfnamefont {T.}~\bibnamefont {Neupert}},\ }\bibfield  {title} {\bibinfo
  {title} {Higher-order topological insulators},\ }\bibfield  {journal}
  {\bibinfo  {journal} {Science Advances}\ }\textbf {\bibinfo {volume} {4}},\
  \href {https://doi.org/10.1126/sciadv.aat0346} {10.1126/sciadv.aat0346}
  (\bibinfo {year} {2018}),\ \Eprint
  {https://arxiv.org/abs/https://advances.sciencemag.org/content/4/6/eaat0346.full.pdf}
  {https://advances.sciencemag.org/content/4/6/eaat0346.full.pdf} \BibitemShut
  {NoStop}%
\bibitem [{\citenamefont {Landau}(1937)}]{lan:37}%
  \BibitemOpen
  \bibfield  {author} {\bibinfo {author} {\bibfnamefont {L.~D.}\ \bibnamefont
  {Landau}},\ }\bibfield  {title} {\bibinfo {title} {On the theory of phase
  transitions. i.},\ }\href@noop {} {\bibfield  {journal} {\bibinfo  {journal}
  {Zh. Eksp. Teor. Fiz.}\ }\textbf {\bibinfo {volume} {7}},\ \bibinfo {pages}
  {19} (\bibinfo {year} {1937})}\BibitemShut {NoStop}%
\bibitem [{\citenamefont {Zanardi}\ and\ \citenamefont
  {Paunkovi{\'c}}(2006)}]{zan:pau:06}%
  \BibitemOpen
  \bibfield  {author} {\bibinfo {author} {\bibfnamefont {P.}~\bibnamefont
  {Zanardi}}\ and\ \bibinfo {author} {\bibfnamefont {N.}~\bibnamefont
  {Paunkovi{\'c}}},\ }\bibfield  {title} {\bibinfo {title} {Ground state
  overlap and quantum phase transitions},\ }\href@noop {} {\bibfield  {journal}
  {\bibinfo  {journal} {Physical Review E}\ }\textbf {\bibinfo {volume} {74}},\
  \bibinfo {pages} {031123} (\bibinfo {year} {2006})}\BibitemShut {NoStop}%
\bibitem [{\citenamefont {Paunkovi{\'c}}\ \emph {et~al.}(2008)\citenamefont
  {Paunkovi{\'c}}, \citenamefont {Sacramento}, \citenamefont {Nogueira},
  \citenamefont {Vieira},\ and\ \citenamefont
  {Dugaev}}]{pau:sac:nog:vie:dug:08}%
  \BibitemOpen
  \bibfield  {author} {\bibinfo {author} {\bibfnamefont {N.}~\bibnamefont
  {Paunkovi{\'c}}}, \bibinfo {author} {\bibfnamefont {P.}~\bibnamefont
  {Sacramento}}, \bibinfo {author} {\bibfnamefont {P.}~\bibnamefont
  {Nogueira}}, \bibinfo {author} {\bibfnamefont {V.}~\bibnamefont {Vieira}},\
  and\ \bibinfo {author} {\bibfnamefont {V.}~\bibnamefont {Dugaev}},\
  }\bibfield  {title} {\bibinfo {title} {Fidelity between partial states as a
  signature of quantum phase transitions},\ }\href@noop {} {\bibfield
  {journal} {\bibinfo  {journal} {Physical Review A}\ }\textbf {\bibinfo
  {volume} {77}},\ \bibinfo {pages} {052302} (\bibinfo {year}
  {2008})}\BibitemShut {NoStop}%
\bibitem [{\citenamefont {Paunkovi{\'c}}\ and\ \citenamefont
  {Vieira}(2008)}]{pau:vie:08}%
  \BibitemOpen
  \bibfield  {author} {\bibinfo {author} {\bibfnamefont {N.}~\bibnamefont
  {Paunkovi{\'c}}}\ and\ \bibinfo {author} {\bibfnamefont {V.~R.}\ \bibnamefont
  {Vieira}},\ }\bibfield  {title} {\bibinfo {title} {Macroscopic
  distinguishability between quantum states defining different phases of
  matter: Fidelity and the {U}hlmann geometric phase},\ }\href@noop {}
  {\bibfield  {journal} {\bibinfo  {journal} {Physical Review E}\ }\textbf
  {\bibinfo {volume} {77}},\ \bibinfo {pages} {011129} (\bibinfo {year}
  {2008})}\BibitemShut {NoStop}%
\bibitem [{\citenamefont {Zanardi}\ \emph
  {et~al.}(2007{\natexlab{a}})\citenamefont {Zanardi}, \citenamefont {Venuti},\
  and\ \citenamefont {Giorda}}]{zan:ven:cam:gio:07}%
  \BibitemOpen
  \bibfield  {author} {\bibinfo {author} {\bibfnamefont {P.}~\bibnamefont
  {Zanardi}}, \bibinfo {author} {\bibfnamefont {L.~C.}\ \bibnamefont
  {Venuti}},\ and\ \bibinfo {author} {\bibfnamefont {P.}~\bibnamefont
  {Giorda}},\ }\bibfield  {title} {\bibinfo {title} {Bures metric over thermal
  state manifolds and quantum criticality},\ }\href@noop {} {\bibfield
  {journal} {\bibinfo  {journal} {Physical Review A}\ }\textbf {\bibinfo
  {volume} {76}},\ \bibinfo {pages} {062318} (\bibinfo {year}
  {2007}{\natexlab{a}})}\BibitemShut {NoStop}%
\bibitem [{\citenamefont {Zanardi}\ \emph
  {et~al.}(2007{\natexlab{b}})\citenamefont {Zanardi}, \citenamefont {Giorda},\
  and\ \citenamefont {Cozzini}}]{zan:goo:coz:07}%
  \BibitemOpen
  \bibfield  {author} {\bibinfo {author} {\bibfnamefont {P.}~\bibnamefont
  {Zanardi}}, \bibinfo {author} {\bibfnamefont {P.}~\bibnamefont {Giorda}},\
  and\ \bibinfo {author} {\bibfnamefont {M.}~\bibnamefont {Cozzini}},\
  }\bibfield  {title} {\bibinfo {title} {Information-theoretic differential
  geometry of quantum phase transitions},\ }\href
  {https://doi.org/10.1103/PhysRevLett.99.100603} {\bibfield  {journal}
  {\bibinfo  {journal} {Phys. Rev. Lett.}\ }\textbf {\bibinfo {volume} {99}},\
  \bibinfo {pages} {100603} (\bibinfo {year} {2007}{\natexlab{b}})}\BibitemShut
  {NoStop}%
\bibitem [{\citenamefont {Campos~Venuti}\ and\ \citenamefont
  {Zanardi}(2007)}]{ven:zan:07}%
  \BibitemOpen
  \bibfield  {author} {\bibinfo {author} {\bibfnamefont {L.}~\bibnamefont
  {Campos~Venuti}}\ and\ \bibinfo {author} {\bibfnamefont {P.}~\bibnamefont
  {Zanardi}},\ }\bibfield  {title} {\bibinfo {title} {Quantum critical scaling
  of the geometric tensors},\ }\href
  {https://doi.org/10.1103/PhysRevLett.99.095701} {\bibfield  {journal}
  {\bibinfo  {journal} {Phys. Rev. Lett.}\ }\textbf {\bibinfo {volume} {99}},\
  \bibinfo {pages} {095701} (\bibinfo {year} {2007})}\BibitemShut {NoStop}%
\bibitem [{\citenamefont {You}\ \emph {et~al.}(2007)\citenamefont {You},
  \citenamefont {Li},\ and\ \citenamefont {Gu}}]{you:wai:gu:07}%
  \BibitemOpen
  \bibfield  {author} {\bibinfo {author} {\bibfnamefont {W.-L.}\ \bibnamefont
  {You}}, \bibinfo {author} {\bibfnamefont {Y.-W.}\ \bibnamefont {Li}},\ and\
  \bibinfo {author} {\bibfnamefont {S.-J.}\ \bibnamefont {Gu}},\ }\bibfield
  {title} {\bibinfo {title} {Fidelity, dynamic structure factor, and
  susceptibility in critical phenomena},\ }\href
  {https://doi.org/10.1103/PhysRevE.76.022101} {\bibfield  {journal} {\bibinfo
  {journal} {Phys. Rev. E}\ }\textbf {\bibinfo {volume} {76}},\ \bibinfo
  {pages} {022101} (\bibinfo {year} {2007})}\BibitemShut {NoStop}%
\bibitem [{\citenamefont {Carollo}\ \emph {et~al.}(2018)\citenamefont
  {Carollo}, \citenamefont {Spagnolo},\ and\ \citenamefont
  {Valenti}}]{car:spa:val:18}%
  \BibitemOpen
  \bibfield  {author} {\bibinfo {author} {\bibfnamefont {A.}~\bibnamefont
  {Carollo}}, \bibinfo {author} {\bibfnamefont {B.}~\bibnamefont {Spagnolo}},\
  and\ \bibinfo {author} {\bibfnamefont {D.}~\bibnamefont {Valenti}},\
  }\bibfield  {title} {\bibinfo {title} {Uhlmann curvature in dissipative phase
  transitions},\ }\href@noop {} {\bibfield  {journal} {\bibinfo  {journal}
  {Scientific reports}\ }\textbf {\bibinfo {volume} {8}},\ \bibinfo {pages} {1}
  (\bibinfo {year} {2018})}\BibitemShut {NoStop}%
\bibitem [{\citenamefont {Ozawa}\ and\ \citenamefont
  {Goldman}(2018)}]{oza:gol:18}%
  \BibitemOpen
  \bibfield  {author} {\bibinfo {author} {\bibfnamefont {T.}~\bibnamefont
  {Ozawa}}\ and\ \bibinfo {author} {\bibfnamefont {N.}~\bibnamefont
  {Goldman}},\ }\bibfield  {title} {\bibinfo {title} {Extracting the quantum
  metric tensor through periodic driving},\ }\href
  {https://doi.org/10.1103/PhysRevB.97.201117} {\bibfield  {journal} {\bibinfo
  {journal} {Phys. Rev. B}\ }\textbf {\bibinfo {volume} {97}},\ \bibinfo
  {pages} {201117} (\bibinfo {year} {2018})}\BibitemShut {NoStop}%
\bibitem [{\citenamefont {Leonforte}\ \emph {et~al.}(2019)\citenamefont
  {Leonforte}, \citenamefont {Valenti}, \citenamefont {Spagnolo},\ and\
  \citenamefont {Carollo}}]{leo:val:spa:car:19}%
  \BibitemOpen
  \bibfield  {author} {\bibinfo {author} {\bibfnamefont {L.}~\bibnamefont
  {Leonforte}}, \bibinfo {author} {\bibfnamefont {D.}~\bibnamefont {Valenti}},
  \bibinfo {author} {\bibfnamefont {B.}~\bibnamefont {Spagnolo}},\ and\
  \bibinfo {author} {\bibfnamefont {A.}~\bibnamefont {Carollo}},\ }\bibfield
  {title} {\bibinfo {title} {Uhlmann number in translational invariant
  systems},\ }\href@noop {} {\bibfield  {journal} {\bibinfo  {journal}
  {Scientific reports}\ }\textbf {\bibinfo {volume} {9}},\ \bibinfo {pages} {1}
  (\bibinfo {year} {2019})}\BibitemShut {NoStop}%
\bibitem [{\citenamefont {Carollo}\ \emph {et~al.}(2020)\citenamefont
  {Carollo}, \citenamefont {Valenti},\ and\ \citenamefont
  {Spagnolo}}]{car:val:spag:20}%
  \BibitemOpen
  \bibfield  {author} {\bibinfo {author} {\bibfnamefont {A.}~\bibnamefont
  {Carollo}}, \bibinfo {author} {\bibfnamefont {D.}~\bibnamefont {Valenti}},\
  and\ \bibinfo {author} {\bibfnamefont {B.}~\bibnamefont {Spagnolo}},\
  }\bibfield  {title} {\bibinfo {title} {Geometry of quantum phase
  transitions},\ }\href@noop {} {\bibfield  {journal} {\bibinfo  {journal}
  {Physics Reports}\ }\textbf {\bibinfo {volume} {838}},\ \bibinfo {pages} {1}
  (\bibinfo {year} {2020})}\BibitemShut {NoStop}%
\bibitem [{\citenamefont {Kitaev}(2001)}]{kit:01}%
  \BibitemOpen
  \bibfield  {author} {\bibinfo {author} {\bibfnamefont {A.~Y.}\ \bibnamefont
  {Kitaev}},\ }\bibfield  {title} {\bibinfo {title} {Unpaired majorana fermions
  in quantum wires},\ }\href@noop {} {\bibfield  {journal} {\bibinfo  {journal}
  {Physics-Uspekhi}\ }\textbf {\bibinfo {volume} {44}},\ \bibinfo {pages} {131}
  (\bibinfo {year} {2001})}\BibitemShut {NoStop}%
\bibitem [{\citenamefont {Mera}\ \emph
  {et~al.}(2017{\natexlab{a}})\citenamefont {Mera}, \citenamefont {Vlachou},
  \citenamefont {Paunkovi{\'c}},\ and\ \citenamefont
  {Vieira}}]{mer:vla:pau:vi:17}%
  \BibitemOpen
  \bibfield  {author} {\bibinfo {author} {\bibfnamefont {B.}~\bibnamefont
  {Mera}}, \bibinfo {author} {\bibfnamefont {C.}~\bibnamefont {Vlachou}},
  \bibinfo {author} {\bibfnamefont {N.}~\bibnamefont {Paunkovi{\'c}}},\ and\
  \bibinfo {author} {\bibfnamefont {V.~R.}\ \bibnamefont {Vieira}},\ }\bibfield
   {title} {\bibinfo {title} {Uhlmann connection in fermionic systems
  undergoing phase transitions},\ }\href@noop {} {\bibfield  {journal}
  {\bibinfo  {journal} {Physical Review Letters}\ }\textbf {\bibinfo {volume}
  {119}},\ \bibinfo {pages} {015702} (\bibinfo {year}
  {2017}{\natexlab{a}})}\BibitemShut {NoStop}%
\bibitem [{\citenamefont {Mera}\ \emph
  {et~al.}(2017{\natexlab{b}})\citenamefont {Mera}, \citenamefont {Vlachou},
  \citenamefont {Paunkovi{\'c}},\ and\ \citenamefont
  {Vieira}}]{mer:vla:pau:vie:17:qw}%
  \BibitemOpen
  \bibfield  {author} {\bibinfo {author} {\bibfnamefont {B.}~\bibnamefont
  {Mera}}, \bibinfo {author} {\bibfnamefont {C.}~\bibnamefont {Vlachou}},
  \bibinfo {author} {\bibfnamefont {N.}~\bibnamefont {Paunkovi{\'c}}},\ and\
  \bibinfo {author} {\bibfnamefont {V.~R.}\ \bibnamefont {Vieira}},\ }\bibfield
   {title} {\bibinfo {title} {Boltzmann--{G}ibbs states in topological quantum
  walks and associated many-body systems: fidelity and {U}hlmann parallel
  transport analysis of phase transitions},\ }\href@noop {} {\bibfield
  {journal} {\bibinfo  {journal} {Journal of Physics A: Mathematical and
  Theoretical}\ }\textbf {\bibinfo {volume} {50}},\ \bibinfo {pages} {365302}
  (\bibinfo {year} {2017}{\natexlab{b}})}\BibitemShut {NoStop}%
\bibitem [{\citenamefont {Amin}\ \emph {et~al.}(2018)\citenamefont {Amin},
  \citenamefont {Mera}, \citenamefont {Vlachou}, \citenamefont {Paunkovi\'c},\
  and\ \citenamefont {Vieira}}]{ami:mer:vla:pau:vie:18}%
  \BibitemOpen
  \bibfield  {author} {\bibinfo {author} {\bibfnamefont {S.~T.}\ \bibnamefont
  {Amin}}, \bibinfo {author} {\bibfnamefont {B.}~\bibnamefont {Mera}}, \bibinfo
  {author} {\bibfnamefont {C.}~\bibnamefont {Vlachou}}, \bibinfo {author}
  {\bibfnamefont {N.}~\bibnamefont {Paunkovi\'c}},\ and\ \bibinfo {author}
  {\bibfnamefont {V.~R.}\ \bibnamefont {Vieira}},\ }\bibfield  {title}
  {\bibinfo {title} {Fidelity and {U}hlmann connection analysis of topological
  phase transitions in two dimensions},\ }\href
  {https://doi.org/10.1103/PhysRevB.98.245141} {\bibfield  {journal} {\bibinfo
  {journal} {Phys. Rev. B}\ }\textbf {\bibinfo {volume} {98}},\ \bibinfo
  {pages} {245141} (\bibinfo {year} {2018})}\BibitemShut {NoStop}%
\bibitem [{\citenamefont {Sacramento}\ \emph {et~al.}(2019)\citenamefont
  {Sacramento}, \citenamefont {Mera},\ and\ \citenamefont
  {Paunkovi{\'c}}}]{sac:mer:pau:19}%
  \BibitemOpen
  \bibfield  {author} {\bibinfo {author} {\bibfnamefont {P.}~\bibnamefont
  {Sacramento}}, \bibinfo {author} {\bibfnamefont {B.}~\bibnamefont {Mera}},\
  and\ \bibinfo {author} {\bibfnamefont {N.}~\bibnamefont {Paunkovi{\'c}}},\
  }\bibfield  {title} {\bibinfo {title} {Vanishing k-space fidelity and phase
  diagram’s bulk--edge--bulk correspondence},\ }\href@noop {} {\bibfield
  {journal} {\bibinfo  {journal} {Annals of Physics}\ }\textbf {\bibinfo
  {volume} {401}},\ \bibinfo {pages} {40} (\bibinfo {year} {2019})}\BibitemShut
  {NoStop}%
\bibitem [{\citenamefont {Amin}\ \emph {et~al.}(2019)\citenamefont {Amin},
  \citenamefont {Mera}, \citenamefont {Paunkovi{\'{c}}},\ and\ \citenamefont
  {Vieira}}]{ami:mer:pau:vie:19}%
  \BibitemOpen
  \bibfield  {author} {\bibinfo {author} {\bibfnamefont {S.~T.}\ \bibnamefont
  {Amin}}, \bibinfo {author} {\bibfnamefont {B.}~\bibnamefont {Mera}}, \bibinfo
  {author} {\bibfnamefont {N.}~\bibnamefont {Paunkovi{\'{c}}}},\ and\ \bibinfo
  {author} {\bibfnamefont {V.~R.}\ \bibnamefont {Vieira}},\ }\bibfield  {title}
  {\bibinfo {title} {Information geometric analysis of long range topological
  superconductors},\ }\href {https://doi.org/10.1088/1361-648x/ab3c70}
  {\bibfield  {journal} {\bibinfo  {journal} {Journal of Physics: Condensed
  Matter}\ }\textbf {\bibinfo {volume} {31}},\ \bibinfo {pages} {485402}
  (\bibinfo {year} {2019})}\BibitemShut {NoStop}%
\bibitem [{\citenamefont {Bhattacharya}\ \emph {et~al.}(2017)\citenamefont
  {Bhattacharya}, \citenamefont {Bandyopadhyay},\ and\ \citenamefont
  {Dutta}}]{bha:ban:dut:17}%
  \BibitemOpen
  \bibfield  {author} {\bibinfo {author} {\bibfnamefont {U.}~\bibnamefont
  {Bhattacharya}}, \bibinfo {author} {\bibfnamefont {S.}~\bibnamefont
  {Bandyopadhyay}},\ and\ \bibinfo {author} {\bibfnamefont {A.}~\bibnamefont
  {Dutta}},\ }\bibfield  {title} {\bibinfo {title} {Mixed state dynamical
  quantum phase transitions},\ }\href@noop {} {\bibfield  {journal} {\bibinfo
  {journal} {Physical Review B}\ }\textbf {\bibinfo {volume} {96}},\ \bibinfo
  {pages} {180303} (\bibinfo {year} {2017})}\BibitemShut {NoStop}%
\bibitem [{\citenamefont {Mera}\ \emph {et~al.}(2018)\citenamefont {Mera},
  \citenamefont {Vlachou}, \citenamefont {Paunkovi\'c}, \citenamefont
  {Vieira},\ and\ \citenamefont {Viyuela}}]{mer:vla:pau:vie:viy:18}%
  \BibitemOpen
  \bibfield  {author} {\bibinfo {author} {\bibfnamefont {B.}~\bibnamefont
  {Mera}}, \bibinfo {author} {\bibfnamefont {C.}~\bibnamefont {Vlachou}},
  \bibinfo {author} {\bibfnamefont {N.}~\bibnamefont {Paunkovi\'c}}, \bibinfo
  {author} {\bibfnamefont {V.~R.}\ \bibnamefont {Vieira}},\ and\ \bibinfo
  {author} {\bibfnamefont {O.}~\bibnamefont {Viyuela}},\ }\bibfield  {title}
  {\bibinfo {title} {Dynamical phase transitions at finite temperature from
  fidelity and interferometric loschmidt echo induced metrics},\ }\href
  {https://doi.org/10.1103/PhysRevB.97.094110} {\bibfield  {journal} {\bibinfo
  {journal} {Phys. Rev. B}\ }\textbf {\bibinfo {volume} {97}},\ \bibinfo
  {pages} {094110} (\bibinfo {year} {2018})}\BibitemShut {NoStop}%
\bibitem [{\citenamefont {Bandyopadhyay}\ and\ \citenamefont
  {Dutta}(2020{\natexlab{a}})}]{ban:dut:20a}%
  \BibitemOpen
  \bibfield  {author} {\bibinfo {author} {\bibfnamefont {S.}~\bibnamefont
  {Bandyopadhyay}}\ and\ \bibinfo {author} {\bibfnamefont {A.}~\bibnamefont
  {Dutta}},\ }\bibfield  {title} {\bibinfo {title} {Unitary preparation of
  many-body chern insulators: Adiabatic bulk-boundary correspondence},\
  }\href@noop {} {\bibfield  {journal} {\bibinfo  {journal} {Physical Review
  B}\ }\textbf {\bibinfo {volume} {102}},\ \bibinfo {pages} {094301} (\bibinfo
  {year} {2020}{\natexlab{a}})}\BibitemShut {NoStop}%
\bibitem [{\citenamefont {Bandyopadhyay}\ and\ \citenamefont
  {Dutta}(2020{\natexlab{b}})}]{ban:dut:20b}%
  \BibitemOpen
  \bibfield  {author} {\bibinfo {author} {\bibfnamefont {S.}~\bibnamefont
  {Bandyopadhyay}}\ and\ \bibinfo {author} {\bibfnamefont {A.}~\bibnamefont
  {Dutta}},\ }\bibfield  {title} {\bibinfo {title} {Dissipative preparation of
  many-body floquet chern insulators},\ }\href@noop {} {\bibfield  {journal}
  {\bibinfo  {journal} {Physical Review B}\ }\textbf {\bibinfo {volume}
  {102}},\ \bibinfo {pages} {184302} (\bibinfo {year}
  {2020}{\natexlab{b}})}\BibitemShut {NoStop}%
\bibitem [{\citenamefont {Sj\"oqvist}\ \emph {et~al.}(2000)\citenamefont
  {Sj\"oqvist}, \citenamefont {Pati}, \citenamefont {Ekert}, \citenamefont
  {Anandan}, \citenamefont {Ericsson}, \citenamefont {Oi},\ and\ \citenamefont
  {Vedral}}]{sjo:00}%
  \BibitemOpen
  \bibfield  {author} {\bibinfo {author} {\bibfnamefont {E.}~\bibnamefont
  {Sj\"oqvist}}, \bibinfo {author} {\bibfnamefont {A.~K.}\ \bibnamefont
  {Pati}}, \bibinfo {author} {\bibfnamefont {A.}~\bibnamefont {Ekert}},
  \bibinfo {author} {\bibfnamefont {J.~S.}\ \bibnamefont {Anandan}}, \bibinfo
  {author} {\bibfnamefont {M.}~\bibnamefont {Ericsson}}, \bibinfo {author}
  {\bibfnamefont {D.~K.~L.}\ \bibnamefont {Oi}},\ and\ \bibinfo {author}
  {\bibfnamefont {V.}~\bibnamefont {Vedral}},\ }\bibfield  {title} {\bibinfo
  {title} {Geometric phases for mixed states in interferometry},\ }\href
  {https://doi.org/10.1103/PhysRevLett.85.2845} {\bibfield  {journal} {\bibinfo
   {journal} {Phys. Rev. Lett.}\ }\textbf {\bibinfo {volume} {85}},\ \bibinfo
  {pages} {2845} (\bibinfo {year} {2000})}\BibitemShut {NoStop}%
\bibitem [{\citenamefont {Sj\"oqvist}(2020)}]{sjo:20}%
  \BibitemOpen
  \bibfield  {author} {\bibinfo {author} {\bibfnamefont {E.}~\bibnamefont
  {Sj\"oqvist}},\ }\bibfield  {title} {\bibinfo {title} {Geometry along
  evolution of mixed quantum states},\ }\href
  {https://doi.org/10.1103/PhysRevResearch.2.013344} {\bibfield  {journal}
  {\bibinfo  {journal} {Phys. Rev. Research}\ }\textbf {\bibinfo {volume}
  {2}},\ \bibinfo {pages} {013344} (\bibinfo {year} {2020})}\BibitemShut
  {NoStop}%
\bibitem [{\citenamefont {Uhlmann}(1986)}]{uhl:86}%
  \BibitemOpen
  \bibfield  {author} {\bibinfo {author} {\bibfnamefont {A.}~\bibnamefont
  {Uhlmann}},\ }\bibfield  {title} {\bibinfo {title} {Parallel transport and
  “quantum holonomy” along density operators},\ }\href@noop {} {\bibfield
  {journal} {\bibinfo  {journal} {Reports on Mathematical Physics}\ }\textbf
  {\bibinfo {volume} {24}},\ \bibinfo {pages} {229} (\bibinfo {year}
  {1986})}\BibitemShut {NoStop}%
\bibitem [{\citenamefont {Matsuura}\ and\ \citenamefont
  {Ryu}(2010)}]{mat:ryu:10}%
  \BibitemOpen
  \bibfield  {author} {\bibinfo {author} {\bibfnamefont {S.}~\bibnamefont
  {Matsuura}}\ and\ \bibinfo {author} {\bibfnamefont {S.}~\bibnamefont {Ryu}},\
  }\bibfield  {title} {\bibinfo {title} {Momentum space metric, nonlocal
  operator, and topological insulators},\ }\href
  {https://doi.org/10.1103/PhysRevB.82.245113} {\bibfield  {journal} {\bibinfo
  {journal} {Phys. Rev. B}\ }\textbf {\bibinfo {volume} {82}},\ \bibinfo
  {pages} {245113} (\bibinfo {year} {2010})}\BibitemShut {NoStop}%
\bibitem [{\citenamefont {Bardyn}\ \emph {et~al.}(2018)\citenamefont {Bardyn},
  \citenamefont {Wawer}, \citenamefont {Altland}, \citenamefont
  {Fleischhauer},\ and\ \citenamefont {Diehl}}]{bar:waw:alt:fle:die:18}%
  \BibitemOpen
  \bibfield  {author} {\bibinfo {author} {\bibfnamefont {C.-E.}\ \bibnamefont
  {Bardyn}}, \bibinfo {author} {\bibfnamefont {L.}~\bibnamefont {Wawer}},
  \bibinfo {author} {\bibfnamefont {A.}~\bibnamefont {Altland}}, \bibinfo
  {author} {\bibfnamefont {M.}~\bibnamefont {Fleischhauer}},\ and\ \bibinfo
  {author} {\bibfnamefont {S.}~\bibnamefont {Diehl}},\ }\bibfield  {title}
  {\bibinfo {title} {Probing the topology of density matrices},\ }\href
  {https://doi.org/10.1103/PhysRevX.8.011035} {\bibfield  {journal} {\bibinfo
  {journal} {Phys. Rev. X}\ }\textbf {\bibinfo {volume} {8}},\ \bibinfo {pages}
  {011035} (\bibinfo {year} {2018})}\BibitemShut {NoStop}%
\bibitem [{\citenamefont {Amin}\ \emph {et~al.}(2020)\citenamefont {Amin},
  \citenamefont {Mera}, \citenamefont {Paunkovi\'c},\ and\ \citenamefont
  {Vieira}}]{ami:mer:pau:vie:20}%
  \BibitemOpen
  \bibfield  {author} {\bibinfo {author} {\bibfnamefont {S.~T.}\ \bibnamefont
  {Amin}}, \bibinfo {author} {\bibfnamefont {B.}~\bibnamefont {Mera}}, \bibinfo
  {author} {\bibfnamefont {N.}~\bibnamefont {Paunkovi\'c}},\ and\ \bibinfo
  {author} {\bibfnamefont {V.~R.}\ \bibnamefont {Vieira}}} (\bibinfo {year} {In
  preparation, 2020})\BibitemShut {NoStop}%
\bibitem [{\citenamefont {Morita}(2001)}]{mor:01}%
  \BibitemOpen
  \bibfield  {author} {\bibinfo {author} {\bibfnamefont {S.}~\bibnamefont
  {Morita}},\ }\href@noop {} {\emph {\bibinfo {title} {Geometry of differential
  forms}}},\ \bibinfo {number} {201}\ (\bibinfo  {publisher} {American
  Mathematical Soc.},\ \bibinfo {year} {2001})\BibitemShut {NoStop}%
\bibitem [{Note1()}]{Note1}%
  \BibitemOpen
  \bibinfo {note} {To see this, observe that a complex matrix can be split into
  its Hermitian and anti-Hermitian components: $Z = Z^H + Z^{AH}$, where $Z^H =
  \protect \frac {1}{2}(Z + Z^\dagger )$ and $Z^{AH} = \protect \frac {1}{2}(Z
  - Z^\dagger )$. This real-linear decomposition divides the full matrix into
  two orthogonal components. Indeed, $\mathop {\protect \rm Re}\nolimits
  \mathop {\protect \rm Tr}\nolimits \left [\left (Z_{1}^{A H}\right )^{\dagger
  } Z_{2}^{H}\right ]=\displaystyle {\protect \frac {1}{2}}\left \{\mathop
  {\protect \rm Tr}\nolimits \left [\left (Z_{1}^{A H}\right )^{\dagger }
  Z_{2}^{H}\right ]+\protect \operatorname {Tr}\left [\left (Z_{2}^{H}\right
  )^{\dagger } Z_{1}^{A H}\right ]\right \}=\displaystyle {\protect \frac
  {1}{2}}\left \{-\protect \operatorname {Tr}\left [\left (Z_{1}^{A H}\right )
  Z_{2}^{H}\right ]+\protect \operatorname {Tr}\left [\left (Z_{2}^{H}\right )
  Z_{1}^{A H}\right ]\right \}= 0$. Moreover, since the real vector space of
  Hermitian matrices and anti-Hermitian matrices both have dimension $k\times
  k$, we conclude that if a complex matrix is (real-)orthogonal to an
  anti-Hermitian matrix, then it must be Hermitian.}\BibitemShut {Stop}%
\end{thebibliography}%
\end{document}